\documentclass[12pt]{article}
\usepackage[utf8]{inputenc}
\usepackage{amsfonts}
\usepackage{latexsym}
\usepackage{amsmath}
\usepackage{amssymb}
\usepackage{amssymb}
\usepackage{slashed}
\usepackage{cancel}
\usepackage{ dsfont }
\usepackage{mathtools}
\usepackage{mathrsfs}
\usepackage{amsthm}
\usepackage[nosort]{cite}
\usepackage{accents}
\usepackage{multicol}
\usepackage{environ}
\usepackage{xcolor}
\usepackage{cite}
\usepackage{enumerate}
\usepackage{alphalph}
\usepackage{etoolbox}
\usepackage[utf8]{inputenc}
\usepackage{pstricks}
\usepackage{longtable}
\usepackage{placeins}
\usepackage{multirow}

\allowdisplaybreaks

\usepackage{geometry}
\usepackage[toc,page]{appendix}

\def\Az{\accentset{(\mathbf{0})}{\mathcal{A}}}
\def\Au{\accentset{(\mathbf{1})}{\mathcal{A}}}
\def\Ad{\accentset{(\mathbf{2})}{\mathcal{A}}}
\def\At{\accentset{(\mathbf{3})}{\mathcal{A}}}

\DeclareMathAlphabet{\mathpzc}{OT1}{pzc}{m}{it}

\usepackage[linktocpage=true]{hyperref}
\hypersetup{
bookmarks=false,         
    unicode=false,          
    pdftoolbar=true,        
    pdfmenubar=true,        
    pdffitwindow=false,     
    pdfstartview={FitH},    
    pdftitle={},    
    pdfauthor={Andrea Fontanella and Luca Romano},     
    pdfsubject={},   
colorlinks=false,
citecolor= red,
linkcolor= red,
urlcolor= red,
filecolor=red,
linkbordercolor={red},
citebordercolor={red},
urlbordercolor={red},
breaklinks=true
}

\def\chapterautorefname~#1\null{Chap.~(#1)\null}
\def\sectionautorefname~#1\null{section~(#1)\null}
\def\subsectionautorefname~#1\null{section~(#1)\null}
\def\figureautorefname~#1\null{Fig.~(#1)\null}
\def\tableautorefname~#1\null{Tab.~(#1)\null}
\def\equationautorefname~#1\null{(#1)\null}
\def\subsubsectionautorefname~#1\null{section~(#1)\null}

\def\appendix#1{\addtocounter{section}{1}\setcounter{equation}{0}
\renewcommand{\thesection}{\Alph{section}}
\section*{Appendix \thesection\protect\indent \parbox[t]{11.15cm}{#1}}
\addcontentsline{toc}{section}{Appendix \thesection\ \ \ #1}}

\usepackage{chngcntr}

\counterwithin*{equation}{section}

\NewEnviron{multieq}[1][2]{
\begin{multicols}{#1}
\begin{subequations}
\setlength{\abovedisplayskip}{-15pt}
\allowdisplaybreaks
\begin{align}
\BODY
\end{align}
\end{subequations}
\end{multicols}
\setlength{\parindent}{0pt}
}

\NewEnviron{multieqB}[1][2]{
\begin{multicols}{#1}
\setlength{\abovedisplayskip}{-15pt}
\allowdisplaybreaks
\begin{align}
\BODY
\end{align}
\end{multicols}
\setlength{\parindent}{0pt}
}

\makeatletter
\newalphalph{\aalphalph}[mult]{\alphalph@alph}{26}
\newcommand{\alphalphval}[1]{%
  \@ifundefined{c@#1}{
    \aalphalph{#1}
  }{%
    \aalphalph{\value{#1}}
  }
}
\makeatother

\AtBeginEnvironment{subequations}{%
}

\begin{document}


\begin{titlepage}
\begin{center}
\vspace*{-1.0cm} 
\hfill HU-EP-20/07 \\

\vspace{2.0cm}

{\LARGE  {\fontfamily{lmodern}\selectfont \bf Lie Algebra Expansion and Integrability\\ [0.4cm]
 in Superstring Sigma-Models}} \\[.2cm]

\vskip 2cm
\textsc{Andrea Fontanella\footnote{\href{mailto:andrea.fontanella@physik.hu-berlin.de}{\texttt{andrea.fontanella@physik.hu-berlin.de}}} \footnotesize and \normalsize Luca Romano\footnote{\href{mailto:lucaromano2607@gmail.com}{\texttt{lucaromano2607@gmail.com}}\ ; address after 01-01-20: Van Swinderen Institute, Groningen University}} 
\\
\vskip 1.2cm

\begin{small}
\textit{$^1$Institut f\"ur Physik, Humboldt-Universit\"at zu Berlin, \\
IRIS Geb\"aude, Zum Grossen Windkanal 6, 12489 Berlin, Germany}

\vspace{5mm}
\textit{$^2$Instituto de F\'isica Te\'orica UAM/CSIC \\
C/ Nicol\'as Cabrera, 13–15, C.U. Cantoblanco, \\
E-28049 Madrid, Spain}
\vspace{5mm}

${}^2${\it Van Swinderen Institute, University of Groningen\\
Nijenborgh 4, 9747 AG Groningen, The Netherlands}\\
\vskip .2truecm

\end{small}

\end{center}

\vskip 1 cm
\begin{abstract}
\vskip1cm
Lie algebra expansion is a technique to generate new Lie algebras from a given one. In this paper, we apply the method of Lie algebra expansion to superstring $\sigma$-models with a $\mathbb{Z}_4$ coset target space. By applying the Lie algebra expansion to the isometry algebra, we obtain different $\sigma$-models, where the number of dynamical fields can change. We reproduce and extend in a systematic way actions of some known string regimes (flat space, BMN and non-relativistic in AdS$_5 \times$S$^5$). 
We define a criterion for the algebra truncation such that the equations of motion of the expanded action of the new $\sigma$-model are equivalent to the vanishing curvature condition of the Lax connection obtained by expanding the Lax connection of the initial model.

\end{abstract}

\end{titlepage}

\tableofcontents
\vspace{5mm}
\hrule


\setcounter{section}{0}

\section*{Introduction}\addcontentsline{toc}{section}{Introduction}

During the last 15 years, since the discovery of integrability in the AdS$_5$/CFT$_4$ duality \cite{Bena:2003wd, Beisert:2010jr, Bombardelli:2016rwb}, a remarkable progress has been made in testing Maldacena's duality \cite{Aharony:1999ti}.
Despite of this, the ambitious long term goal to understand quantum gravity would possibly require a better picture of the \emph{holographic principle}, which generically states that a theory of gravity is equivalent to a gauge theory living on the boundary of the spacetime.
With this in mind, there has been a remarkable effort in investigating the integrable AdS$_d$/CFT$_{d-1}$ duality, with $d <5$, and integrable deformations, \cite{Beisert:2010jr, Bombardelli:2016rwb, Arutyunov:2008if, Stefanski:2008ik, Sorokin:2011rr, Delduc:2013qra}, see also references therein.

Another interesting and fascinating arena where the holographic principle could be tested is provided by taking particular physical limits of the string, such that the final theory shows a different (and possibly simpler) behaviour from the initial one. Some examples are given for instance by the non- and ultra-relativistic limits, i.e. when the speed of light $c$ tends to infinity or zero, respectively \cite{Bergshoeff:2017btm, Hartong:2015xda, Bergshoeff:2014jla, Duval:2014uoa, Cardona:2016ytk, Bergshoeff:2020fiz}. However, limits where other physical parameters of the theory are involved (e.g. radius of some spacetime geometry, string angular momentum) are interesting as well.

During the last decade, aspects of the non-relativistic string have been extensively investigated \cite{Kluson:2018grx, Gomis:2019zyu, Harmark:2019upf, Bergshoeff:2018vfn, Sakaguchi:2007zsa}, due also to interesting applications that Newton-Cartan geometry has found in the context of condensed matter physics and holography \cite{Son:2013rqa, Hoyos:2011ez, Geracie:2014nka, Christensen:2013lma, Christensen:2013rfa, Harmark:2017rpg}.
As an example, in \cite{Gomis:2000bd} it has been proposed a theory described by a $\sigma$-model which has a relativistic worldsheet, and a non-relativistic target space geometry. This theory is unitary, ultraviolet complete and its S-matrix has a non-relativistic symmetry algebra. 
Other  non-relativistic theories have been discussed e.g. in \cite{Gomis:2005pg, Brugues:2004an, Brugues:2006yd}, and aspects such as the equations of motion, T-duality transformations, and the identification of the spacetime geometry, dubbed as string Newton-Cartan geometry and its coupling to matter, have received a growing attention 
\cite{Gomis:2019zyu, Bergshoeff:2018vfn, Bergshoeff:2019pij, Hansen:2020pqs, Yan:2019xsf, Bergshoeff:2018yvt}. 
Classifying non-relativistic spacetime geometries is also an active field of research \cite{Figueroa-OFarrill:2018ilb, Figueroa-OFarrill:2019sex, Figueroa-OFarrill:2019ucc}. 
It turns out that a more systematic and rigorous way to discuss various string regimes is given in the context of the Lie algebra expansion, and this will be discussed extensively in this paper.

The method of Lie algebra expansion, which was rigorously formulated in \cite{deAzcarraga:2002xi, deAzcarraga:2007et} based on the initial work of \cite{Hatsuda:2001pp}, consists in generating new Lie algebras (usually bigger) from a given one, once an initial decomposition is defined. A general feature of the method is that in the lowest order of the expansion, one obtains the \.In\"on\"u-Wigner contraction of the initial algebra with respect to the given decomposition. The contracted algebra gives us information about the physical regime that is inspected by the expansion.

From the mathematical point of view, the idea of Lie algebra expansion consists in rescaling the group coordinates with a parameter $\lambda$. From the physical point of view, this parameter $\lambda$ should be identified with a certain function of some physical parameter(s) appearing in the theory.  One of the powerful features of the Lie algebra expansion is that it comes equipped with the so-called \emph{truncation rules}, which ensure that the expansion is truncated consistently. Recently, the Lie algebra expansion has been applied in the context of the Einstein-Hilbert action \cite{Hansen:2018ofj,Hansen:2019vqf}, later further developed in \cite{Bergshoeff:2019ctr, Hansen:2020pqs}, and also by including supersymmetry \cite{deAzcarraga:2019mdn, Romano:2019ulw}.  Lie algebra expansion has been also further extended to the semigroup expansion method \cite{Izaurieta:2006zz, Salgado:2017goq, Gonzalez:2016xwo, Concha:2019lhn, Concha:2020sjt}.
These methods have recently found interesting applications in several different contexts \cite{Ozdemir:2019tby, Kasikci:2020qsj, Ozdemir:2019orp, Concha:2019dqs, Gomis:2019nih}. 

In this paper we apply the method of Lie algebra expansion to integrable coset $\sigma$-models, which typically appear in the context of string theory, to inspect various physical regimes. The starting point is to write down a 2d action with an associated global and gauge symmetry, $G$ and $H$ respectively, where formally $H \subset G$. The global symmetry algebra Lie$(G)$ is then expanded and appropriately truncated, which in turns implies an expansion of the initial 2d action. 
The action of the initial model is classically integrable.
It is in the aim of this paper to investigate whether classical integrability is also a property of the expanded actions of the new $\sigma$-models\footnote{Classical integrability of the bosonic non-relativistic string has been discussed in \cite{Roychowdhury:2019vzh, Roychowdhury:2019olt, Kluson:2017ufb, Roychowdhury:2020abu}.}.

\vspace{3mm}

\emph{Plan of the paper.} In \autoref{sec:generalities}, we define the 2d string $\sigma$-model and the Lie algebra expansion, and we fix the notation. In \autoref{sec:AdS5xS5}, we apply the Lie algebra expansion to the AdS$_5 \times$S$^5$ supersting action, where we reproduce and extend in a systematic way actions of known regimes (flat space, BMN and non-relativistic). 
In \autoref{sec:integrability}, we give a criterion for the algebra truncation such that the equations of motion of the expanded action of the new $\sigma$-model are equivalent to the zero curvature condition of a Lax connection.  We write the Lax connection for the cases discussed in AdS$_5 \times$S$^5$. 

This paper ends with some appendices, where we give our conventions and the commutation relations of the Lie algebras which appear through the paper.

\section{Lie Algebra Expansion applied to Coset Sigma-Models}
\label{sec:generalities}
\subsection{The Sigma-Model}

In this section we introduce the string $\sigma$-models that will be of central importance in this paper, and we set the notation.
We consider a 2-dimensional string $\sigma$-model with target space $G / H$, where $G$ and $H$ are Lie (super)groups, whose respective Lie (super)algebras are $\mathfrak{g}$ and $\mathfrak{h}$. We assume that $\mathfrak{g}$ admits a $\mathbb{Z}_4$ outer automorphism, such that it decomposes as
\begin{equation}
\mathfrak{g} = \accentset{(\bf 0)}{\mathfrak{g}}\oplus \accentset{(\bf 1)}{\mathfrak{g}}\oplus\accentset{(\bf 2)}{\mathfrak{g}}\oplus\accentset{(\bf 3)}{\mathfrak{g}} \, , \qquad\qquad
 \accentset{(\bf 0)}{\mathfrak{g}} \equiv \mathfrak{h} \, . 
\end{equation}
We denote the string world-sheet coordinates as $\sigma^{\mu} \equiv (\tau , \sigma)$, and the world-sheet metric as $h_{\mu\nu}$. The Levi-Civita symbol is taken with the convention $\epsilon^{\tau\sigma} = - \epsilon_{\tau\sigma} = + 1$. 
The commutation relations of $\mathfrak{g}$ are denoted by 
\begin{equation}
\label{com_relations_g}
[T_A , T_B] = f_{AB}{}^C T_C \, , 
\end{equation}
where $T_A$ are the generators of $\mathfrak{g}$, and we denote by $\accentset{({\bf k})}{T}_A$ the generators of $\accentset{(\bf k)}{\mathfrak{g}}$.
The coset representative $g \in G$ is given by  
\begin{equation}
g(\tau, \sigma) = \exp \bigg(\sum_{{\bf k}=0}^3 \accentset{({\bf k})}{\alpha}^A(\tau , \sigma) \accentset{({\bf k})}{T}_A \bigg) \, ,
\end{equation}
where $\accentset{({\bf k})}{\alpha}^A(\tau , \sigma)$, for ${\bf k} = 1, 2, 3$, are the \emph{fields} carrying the propagating d.o.f. of the $\sigma$-model. 
We introduce the left-invariant Maurer-Cartan 1-form $\mathcal{A}$ given by 
\begin{equation}
\mathcal{A} = - g^{-1} \text{d} g \equiv L^A  T_A\, , 
\end{equation}
which satisfies the Maurer-Cartan equation
\begin{equation}
\label{Maurer_Cartan_eqn1}
\text{d} \mathcal{A} - \frac{1}{2}\mathcal{A} \wedge \mathcal{A} = 0 \, ,
\end{equation}
where the wedge product between Lie algebra valued forms has been defined in \autoref{wedge_prod}. The Maurer-Cartan equation in components reads
\begin{equation}
\label{Maurer_Cartan_eqn2}
\text{d} L^A =  \frac{1}{2} f_{BC}{}^A L^B \wedge L^C \, . 
\end{equation}
We denote Maurer-Cartan 1-forms with calligraphic letters, and Lie algebra generators with latin uppercase letters.  The $\mathbb{Z}_4$ automorphism induces a decomposition of the Maurer-Cartan 1-form as follows
\begin{equation}
\mathcal{A} = \accentset{({\bf 0})}{\mathcal{A}} + \accentset{({\bf 1})}{\mathcal{A}} + \accentset{({\bf 2})}{\mathcal{A}} + \accentset{({\bf 3})}{\mathcal{A}} \, ,
\end{equation}
with 
\begin{align}
\accentset{({\bf k})}{\mathcal{A}} = \accentset{({\bf k})}{L}^A \accentset{({\bf k})}{T}_A \, .
\end{align}
We consider the string $\sigma$-model action 
\begin{equation}
\label{action_master1}
S = - \frac{1}{2} \int_{\partial M_3} \text{d}^2 \sigma  \sqrt{|h|}  h^{\mu\nu} \text{str} ( \accentset{({\bf 2})}{\mathcal{A}}_{\mu} \accentset{({\bf 2})}{\mathcal{A}}_{\nu}) 
- \frac{1}{2} \int_{M_3} \text{str} \Big[ \accentset{({\bf 2})}{\mathcal{A}} \circ
(\accentset{({\bf 3})}{\mathcal{A}} \wedge \accentset{({\bf 3})}{\mathcal{A}}) - \accentset{({\bf 2})}{\mathcal{A}} \circ ( \accentset{({\bf 1})}{\mathcal{A}} \wedge \accentset{({\bf 1})}{\mathcal{A}}) \Big] \, , 
\end{equation}
with the convention defined in \autoref{eq:wedgecirc},  where `str' stands for the inner product\footnote{This cannot be the Killing form, since the latter one must be identically zero in order to make the beta function vanish \cite{Zarembo:2010sg}.} on $\mathfrak{g}$, $M_3$ is a 3-dimensional manifold, whose boundary is the 2-dimensional string world-sheet, and the brackets between the currents in the Wess-Zumino term of \autoref{action_master1} indicate that the (anti)commutator between the generators must be taken. 
The relative coefficient between the kinetic and the Wess-Zumino terms in \autoref{action_master1} is fixed to be $\pm 1$ by requiring invariance of the full action under $\kappa$-symmetry \cite{Arutyunov:2009ga}. Here we choose it to be $+1$.

The equations of motion for the fields are: 
\begin{subequations}
\begin{align}
\label{eq_motion1}
\partial_{\mu} (\sqrt{|h|} h^{\mu\nu} \Ad_{\nu}) - \sqrt{|h|} h^{\mu\nu} [\Az_{\mu} , \Ad_{\nu}] + \frac{1}{2} \epsilon^{\mu\nu} ( [\Au_{\mu} , \Au_{\nu}] - [\At_{\mu}, \At_{\nu}]) =& 0 \ , \\
\label{eq_motion2}
(\sqrt{|h|} h^{\mu\nu} + \epsilon^{\mu\nu})[\At_{\mu}, \Ad_{\nu}] =& 0 \ , \\
\label{eq_motion3}
(\sqrt{|h|} h^{\mu\nu} - \epsilon^{\mu\nu})[\Au_{\mu}, \Ad_{\nu}] =& 0 \ ,
\end{align}
\end{subequations}
It is interesting from the classical integrability point of view of the theory that this set of equations is equivalent to the condition of vanishing curvature 
\begin{align}
d\mathscr{L}-\frac{1}{2}\mathscr{L}\wedge \mathscr{L}=0\ ,
\end{align}
of the following Lax connection \cite{Arutyunov:2009ga} 
\begin{equation}
\label{Lax_connection}
\mathscr{L}_{\mu} = \ell_0 \Az_{\mu} + \ell_1 \Ad_{\mu} + \ell_2 \frac{1}{\sqrt{|h|}}h_{\mu\nu} \epsilon^{\nu\rho}\Ad_{\rho} + \ell_3 \Au_{\mu} + \ell_4 \At_{\mu} \ ,
\end{equation}
where the parameters $\ell_i$ can be parametrised in terms of a single spectral parameter $z$ as 
\begin{equation}
\ell_0 = 1 \, , \quad
\ell_1 = \frac{1}{2} \bigg( z^2 + \frac{1}{z^2} \bigg) \, , \quad
\ell_2 = - \frac{1}{2} \bigg( z^2 - \frac{1}{z^2} \bigg) \, , \quad
\ell_3 = z \, , \quad
\ell_4 = \frac{1}{z} \, .
\end{equation}
Having described the $\sigma$-model and some aspects of its classical integrability, in the next section we shall introduce the Lie algebra expansion method.

\subsection{Lie Algebra Expansion}\label{subsec:liealgbrexpansion}
In this section we give a brief introduction to the Lie algebra expansion method, referring to \cite{deAzcarraga:2002xi, deAzcarraga:2007et, Hatsuda:2001pp} for a more detailed analysis.  This method is a fundamental tool that will give us a general procedure to inspect different physical regimes, e.g. the non-relativistic one, starting from the AdS$_{5}\times$S$^{5}$ string $\sigma$-model.

The expansion procedure for a given Lie superalgebra is a systematic way to generate a new superalgebra, usually bigger that the starting one. The original approach described in \cite{deAzcarraga:2002xi} consists in  decomposing the initial superalgebra in subspaces $V_0, V_1, V_2, ..., V_n$. Such decomposition is associated with a rescaling of the group parameters by a constant $\lambda$. This, in turn, implies that the Maurer-Cartan 1-forms are expanded as power series in $\lambda$. By studying the Maurer-Cartan equations term by term in the expansion, it is possible to define the truncation criteria for the expanded 1-forms, in such a way that the coefficients in the expansion could be seen as the Maurer-Cartan 1-forms of a new algebra. In this way, it is possible to generate a new algebra from the initial one, which is in general bigger. 
For convenience, we will perform the expansion directly at the level of the generators, where every term in the corresponding power series plays the role of a new generator in the new algebra, the structure constant being fixed by the analysis of the Maurer-Cartan equations.

\subsubsection[{Decomposition $V_0 \oplus V_1$}]{Decomposition $\mathbf{V_0} \oplus \mathbf{V_1}$}
\label{Sec:expansion_Z2}
The expansion rules are determined by the decomposition of the algebra, and different decompositions lead to different physical regimes. 
We consider the following decomposition 
\begin{align}
\label{Z2_splitting}
\mathfrak{g}=V_{0}\oplus V_{1}
\end{align}
where $V_0$ and $V_1$ have a \emph{symmetric space structure}, 
\begin{align}
\label{symmetric_space_property}
[V_{0},V_{0}]\subseteq V_{0}\qquad\qquad 
[V_{0},V_{1}]\subseteq V_{1}\qquad\qquad
[V_{1},V_{1}]\subseteq V_{0}\ .
\end{align}
Denoting the generators of each subspaces $V_0$ and $V_1$ by
\begin{equation}
V_0 = \text{span} \{ T_{A_0} \} \, , \qquad\qquad
V_1 = \text{span} \{ T_{A_1} \} \, , 
\end{equation}
where $A_0$ and $A_1$ label the generators in the two subspaces, then the algebra expansion is given by 
\begin{subequations}
\begin{alignat}{2}
T_{A_0}=&\sum_{\substack{k=0,\ k\text{ even}}}^{N_{0}}\lambda^{k}\ \accentset{(k)}{T}_{A_0}&&=\accentset{(0)}{T}_{A_0}+\lambda^2\ \accentset{(2)}{T}_{A_0}+... \ ,\\
T_{A_1} =&\sum_{\substack{k=1,\ k\text{ odd}}}^{N_{1}}\lambda^{k}\ \accentset{(k)}{T}_{A_1} &&=\lambda\ \accentset{(1)}{T}_{A_1} +\lambda^3\ \accentset{(3)}{T}_{A_1}+... \ ,
\end{alignat}
\end{subequations} 
where $N_{0}$ and $N_{1}$ are even and odd natural numbers respectively,  and they represent a truncation of the infinite expansion.
The truncated expanded algebra is called $\mathfrak{g}(N_0, N_1)$, and the truncation is consistent if the following \emph{truncation conditions} are satisfied \cite{deAzcarraga:2002xi} 
\begin{align}
N_{1}=N_{0}\pm 1\ .
\end{align}
If we denote the commutation relations of $\mathfrak{g}$ as in \autoref{com_relations_g}, then the commutation relations between the generators of $\mathfrak{g}(N_0, N_1)$ are
\begin{equation}
\label{expanded_comm_rel}
[ \accentset{(m)}{T}_A , \accentset{(n)}{T}_B ] = f_{AB}{}^C \ \, \accentset{(m+n)}{T}_C \, ,
\end{equation}
which are inherited from \autoref{com_relations_g}, and are zero when the order in $\lambda$ on l.h.s does not match the order in $\lambda$ on r.h.s..  We stress that if the order $m+n$ exceeds the truncation order then the structure constant vanishes. The lowest order of the expansion, which is given by $\mathfrak{g}(0, 1)$, corresponds to the \.In\"on\"u-Wigner contraction of the initial algebra with respect to the subalgebra $V_0$. This algebra gives us the information about which physical regime is inspected in the expansion.\\

The $\mathbb{Z}_4$ grading does not need to be aligned in general with the $V_0 \oplus V_1$ decomposition. This means that any generator $\accentset{(\bf k)}{T}$ belonging to $\accentset{(\bf k)}{\mathfrak{h}}$ will have, in general, a component along both $V_{0}$ and $V_{1}$. To consider this, we decompose the Maurer-Cartan 1-form $\accentset{({\bf k})}{\mathcal{A}}$ as
\begin{align}
\accentset{({\bf k})}{\mathcal{A}}=\accentset{({\bf k})}{\mathcal{B}}+\accentset{({\bf k})}{\mathcal{C}} \, , \qquad
 \accentset{({\bf k})}{\mathcal{B}}\in\ V_{0} \, , \qquad \accentset{({\bf k})}{\mathcal{C}}\in\ V_{1} \, , 
\end{align}
where $\accentset{({\bf k})}{\mathcal{B}}$ and $\accentset{({\bf k})}{\mathcal{C}}$ expand as
\begin{subequations}
\begin{alignat}{2}
\accentset{(\bf k)}{\mathcal{B}}=&\sum_{\substack{i=0,\ i\text{ even}}}^{N_{0}}\lambda^{i}\ \accentset{({\bf k}, i)}{B}&&=\accentset{({\bf k}, 0)}{\mathcal{B}}+\lambda^2\ \accentset{({\bf k},2)}{\mathcal{B}}+... \ ,\\
\accentset{(\bf k)}{\mathcal{C}}=&\sum_{\substack{i=1,\ i\text{ odd}}}^{N_{1}}\lambda^{i}\ \accentset{({\bf k}, i)}{\mathcal{C}}&&=\lambda\ \accentset{({\bf k}, 1)}{\mathcal{C}}+\lambda^3\ \accentset{({\bf k},3)}{\mathcal{C}}+... \ .
\end{alignat}
\end{subequations} 
The reader should be careful to distinguish the bold index ${\bf k}$, which refers to the $\mathbb{Z}_4$ grading, and the index $i$, which refers to the expansion in $\lambda$.  

This procedure induces an expansion of the \emph{components} $\accentset{({\bf k})}{L}^A$ of the Maurer-Cartan 1-form as
\begin{equation}
\accentset{({\bf k})}{L}^A = \accentset{({\bf k})}{M}^A + \accentset{({\bf k})}{N}^A \, , 
\end{equation}
where 
\begin{equation}
\accentset{(\bf k)}{\mathcal{B}} = \accentset{({\bf k})}{M}^{A_0} \accentset{({\bf k})}{T}_{A_0}\, , \qquad\qquad 
\accentset{(\bf k)}{\mathcal{C}} = \accentset{({\bf k})}{N}^{A_1} \accentset{({\bf k})}{T}_{A_1} \, ,
\end{equation}
and the induced expansion is
\begin{subequations}
\begin{alignat}{2}
\accentset{(\bf k)}{M}^{A_0}=&\sum_{\substack{i=0,\ i\text{ even}}}^{N_{0}}\lambda^{i}\ \accentset{({\bf k}, i)}{M}^{A_0}&&=\accentset{({\bf k}, 0)}{M}^{A_0} +\lambda^2\ \accentset{({\bf k},2)}{M}^{A_0}+... \ ,\\
\accentset{(\bf k)}{N}^{A_1} =&\sum_{\substack{i=1,\ i\text{ odd}}}^{N_{1}}\lambda^{i}\ \accentset{({\bf k}, i)}{N}^{A_1}&&=\lambda\ \accentset{({\bf k}, 1)}{N}^{A_1} +\lambda^3\ \accentset{({\bf k},3)}{N}^{A_1}+... \ .
\end{alignat}
\end{subequations} 
At this point, we are interested in the dynamic of the initial $\sigma$-model after performing the Lie algebra expansion described above. In general, the action \autoref{action_master1}, keeping the bilinear product fixed from the beginning, will expand as 
\begin{equation}
\label{action_exp_Z2}
S = \accentset{(0)}{S} + \lambda\  \accentset{(1)}{S} +  ... \, , 
\end{equation}
where $\accentset{(0)}{S}$ is just the action \autoref{action_master1} with all $\accentset{(\bf k)}{L}^A$ replaced by $\accentset{({\bf k}, 0)}{M}^A$, while $\accentset{(1)}{S}$ is the action \autoref{action_master1} where all $\accentset{(\bf k)}{L}^A$ are replaced by $\accentset{({\bf k}, 0)}{M}^A$ except for one of them, which must be replaced by $\accentset{({\bf k}, 1)}{N}^A$, and all possible combinations must be considered. In a similar way one can recursively reconstruct the action $\accentset{(n)}{S}$. The equations of motion  of the generic action $\accentset{(n)}{S}$ and its classical integrability, as well as its global symmetries, will be considered later in \autoref{sec:integrability}.

In this paper we shall always apply the Lie algebra expansion to the string action (\ref{action_master1})  written in terms of the Wess-Zumino 3-form, but not to the equivalent one written in terms of the Wess-Zumino 2-form. As it will be discussed in \autoref{sec:integrability}, the reason for this is because the equations of motion for the generic expanded action $\accentset{(n)}{S}$, obtained from the initial action $S$ written in terms of the Wess-Zumino 3-form, coincide with the initial equations of motion for (\ref{action_master1}) expanded up to the $n$-th order. This is not always the case when the initial action is written in terms of the Wess-Zumino 2-form, and this is a necessary condition in order to apply the techniques presented in this paper. 

In the method of Lie algebra expansion, one should regard all expanded generators as new independent generators of a new Lie algebra. Therefore the $\sigma$-model associated with this new Lie algebra will have in general a bigger number of physical fields, one for each independent generator. The general coset representative of this new $\sigma$-model will be 
\begin{equation}
g(\tau, \sigma) = \exp \Bigg[\sum_{{\bf k}=0}^3 \ \mathop{\sum_{i=0}}_{i\text{ even}}^{N_0} \ \mathop{\sum_{j=1}}_{ j\text{ odd}}^{N_1}\, \bigg(\ \accentset{({\bf k}, i)}{\alpha}^{\ A_0}(\tau , \sigma) \accentset{({\bf k}, i)}{T}_{A_0}  + \accentset{({\bf k}, j)}{\alpha}^{\ A_1}(\tau , \sigma) \accentset{({\bf k}, j)}{T}_{A_1} \bigg)\Bigg] \, ,
\end{equation}
where $\accentset{({\bf k}, i)}{\alpha}^{\ A_0}(\tau , \sigma)$ and $\accentset{({\bf k}, j)}{\alpha}^{\ A_1}(\tau , \sigma)$, for ${\bf k} = 1, 2, 3$, are the fields carrying the propagating d.o.f. of the new $\sigma$-model. 

So far we have introduced the notion of Lie algebra expansion \cite{deAzcarraga:2002xi, Hatsuda:2001pp}. Next, we will apply this method to derive $\sigma$-model actions which describe physics in different regimes in the context of AdS$_5 \times $S$^5$ superstring. 
As we shall see, there are other possible decompositions of $\mathfrak{g}$, such as the three subspace decomposition, which are useful as well. We did not discussed them here in general, but they will be introduced later, tailored for the particular cases considered.

\section{Lie Algebra Expansion in AdS$_5 \times $S$^5$ Superstring}
\label{sec:AdS5xS5}

In the previous section we have introduced the technique of Lie algebra expansion. In this section we shall apply it to investigate different regimes of the AdS$_5 \times $S$^5$ superstring sigma model, drawing a systematic way to reproduce and extend known results.\\ 

The Green-Schwarz action for a string propagating in AdS$_5 \times $S$^5$ has been constructed by Metsaev-Tseytlin \cite{Metsaev:1998it} as a 2-dimensional $\sigma$-model with target space the supercoset
\begin{equation}
G / H_0 \equiv \frac{PSU(2,2|4)}{SO(4,1) \times SO(5)} \ . 
\end{equation} 
We follow the Metsaev-Tseytlin's notation, where the $\mathfrak{psu}(2,2|4)$ superalgebra generators are  
\begin{equation}
P_{\hat{a}}\, ,\quad J_{ab} \, , \quad J_{a'b'}\, , \quad Q_{\alpha\alpha' I} \, , 
\end{equation}
and where the indices run as follows
%

\begin{alignat*}{4}
\hat{a}, \hat{b}, \hat{c}, ... &= 0, ... , 9 && \qquad\qquad \text{AdS$_5 \times$S$^5$ tangent space indices} \\ 
a, b, c, ... &= 0, ... , 4&&\qquad\qquad  \text{AdS$_5$ tangent space indices} \\
a', b', c', ... &= 5, ... , 9 &&\qquad\qquad  \text{S$^5$ tangent space indices} \\
\alpha, \beta, \gamma, ... &= 1, ... , 4 &&\qquad\qquad  \text{AdS$_5$ spinor indices}\\
\alpha', \beta', \gamma', ... &= 1, ... , 4 &&\qquad\qquad  \text{S$^5$ spinor indices}\\
I, J, K, ... &= 1, 2&&\qquad\qquad\text{$\mathcal{N} =2$ supersymmetry indices}
\end{alignat*}
We use the compact notation for the rotation generators $J_{\hat{a}\hat{b}}$, where the constraint $J_{a a'} = 0$ is imposed. We also denote collectively $\hat{\alpha} = (\alpha , \alpha')$.
The $\mathbb{Z}_4$ outer automorphism of $\mathfrak{psu}(2,2|4)$ induces the decomposition
\begin{eqnarray}
\notag
&& \hspace{1cm}\mathfrak{psu}(2, 2|4) = \accentset{({\bf 0})}{\mathfrak{h}} \oplus \accentset{({\bf 1})}{\mathfrak{h}} \oplus \accentset{({\bf 2})}{\mathfrak{h}} \oplus \accentset{({\bf 3})}{\mathfrak{h}} \ , \\
\notag
&& \\
\notag
\accentset{({\bf 0})}{\mathfrak{h}} &=& \mbox{span}\{ J_{\hat{a}\hat{b}}\} \, ,\hspace{2.9cm}
\accentset{({\bf 2})}{\mathfrak{h}} = \mbox{span}\{ P_{\hat{a}} \} \, , \\
\accentset{({\bf 1})}{\mathfrak{h}} &=& \mbox{span}\{Q_{\alpha\alpha' \, 1} \} \, ,  \qquad\qquad\qquad\ 
\accentset{({\bf 3})}{\mathfrak{h}} = \mbox{span}\{Q_{\alpha\alpha' \, 2} \} \ .
\end{eqnarray}
The left-invariant Maurer-Cartan 1-forms are given by
\begin{equation}
\mathcal{A} = L^A T_A = L^{\hat{a}} P_{\hat{a}} + \frac{1}{2} L^{\hat{a}\hat{b}} J_{\hat{a}\hat{b}} + L^{\alpha\alpha' \, I} Q_{\alpha\alpha' \, I} \, ,   
\end{equation}
where one must impose that $L^{a a'} = 0$, as a consequence of $J_{a a'} = 0$.  
The $\mathfrak{psu}(2,2|4)$ superalgebra commutation relations are:
\begin{multieqB}[2]
[P_a,P_b]&=J_{ab}\nonumber\\
[P_{a'},P_{b'}]&=-J_{a'b'}\nonumber\\
[P_{\hat{a}},J_{\hat{b}\hat{c}}]&=\eta_{\hat{a}\hat{b}}P_{\hat{c}}-\eta_{\hat{a}\hat{c}}P_{\hat{b}}\nonumber\\
[J_{\hat{a}\hat{b}},J_{\hat{c}\hat{d}}]&=2 \eta_{\hat{c}[\hat{b}}J_{\hat{a}]\hat{d}} - 2 \eta_{\hat{d}[\hat{b}}J_{\hat{a}]\hat{c}}\nonumber\\
[Q_I,P_{\hat{a}}]&=-\frac{i}{2}\epsilon_{{IJ}}
Q_J\Gamma_{\hat{a}}\nonumber\\
\{Q_{I }, Q_{J}\}&=-2 i \delta_{IJ}
C \Gamma^{\hat{a}} P_{\hat{a}}
+\epsilon_{{IJ}}C \Gamma^{\hat{a}\hat{b}} J_{\hat{a}\hat{b}}\nonumber\\
[Q_I,J_{\hat{a}\hat{b}}]&=-\frac{1}{2} Q_I\Gamma_{\hat{a}\hat{b}}
\end{multieqB}
\noindent The AdS$_5\times $S$^5$ action is the following 
\begin{equation}
S_{\text{AdS}_5\times \text{S}^5} = - \frac{1}{2} \int_{\partial M_3} \text{d}^2 \sigma \sqrt{|h|} h^{\mu\nu} L^{\hat{a}}_{\mu} L^{\hat{b}}_{\nu} \eta_{\hat{a}\hat{b}} - i \int_{M_3} (\tau_3)_{IJ}  \bar{L}^I \wedge L^{\hat{a}} \Gamma_{\hat{a}} \wedge L^J \, , 
\end{equation}
where $\tau_3$ is the third Pauli matrix. Having set the notation for the AdS$_5 \times $S$^5$ $\sigma$-model, now we shall use the Lie algebra expansion to describe its dynamic in various regimes. The regimes considered in this paper are the flat space, BMN and non-relativistic (Newton-Hooke and Galilei).

 \subsection{The Flat Space Case} 
\label{sec:flat_space}
As pointed out in \cite{Metsaev:1998it}, from the AdS$_5 \times$S$^5$ superstring action one can recover the flat space action given in \cite{Henneaux:1984mh}. This is performed by taking the common radius of AdS$_5$ and S$^5$ to be large such that (a subalgebra of) the $\mathcal{N}=2, D =10$ Poincaré superalgebra is obtained as an \.In\"on\"u-Wigner contraction of the $\mathfrak{psu}(2,2|4)$ superalgebra.  However, the two-subspaces decomposition of $\mathfrak{psu}(2,2|4)$ as described in \autoref{subsec:liealgbrexpansion} is not useful in this case, but we need the following three-subspaces decomposition\footnote{The relation between the expansion parameter $\lambda$ and the common radius $R$ is  $\lambda^2 = R$. }
\begin{align}
\mathfrak{psu}(2,2|4) = V_0 \oplus V_1 \oplus V_2  \, , 
\end{align}
where
\begin{align}
&V_{0} = \text{span} \{ J_{\hat{a}\hat{b}} \} \, ,
&&V_{1} = \text{span} \{ Q_{\alpha\alpha' \, I} \} \, , 
&&V_{2} = \text{span} \{ P_{\hat{a}} \}\, . 
\end{align}
A specific analysis of this case leads to the following expansion
\begin{subequations}
\label{3sub_expansion}
\begin{alignat}{2}
J_{\hat{a}\hat{b}}=&\sum_{\substack{k=0,\ k\text{ even}}}^{N_{0}}\lambda^{k}\ \accentset{(k)}{J}_{\hat{a}\hat{b}}&&=\accentset{(0)}{J}_{\hat{a}\hat{b}}+\lambda^2\ \accentset{(2)}{J}_{\hat{a}\hat{b}}+... \ ,\\
Q_{\alpha\alpha' \, I} =&\sum_{\substack{k=1,\ k\text{ odd}}}^{N_{1}}\lambda^{k}\ \accentset{(k)}{Q}_{\alpha\alpha' \, I} &&=\lambda\ \accentset{(1)}{Q}_{\alpha\alpha' \, I} +\lambda^3\ \accentset{(3)}{Q}_{\alpha\alpha' \, I}+... \ , \\
P_{\hat{a}}=&\sum_{\substack{k=2,\ k\text{ even}}}^{N_{2}}\lambda^{k}\ \accentset{(k)}{P}_{\hat{a}}&&=\lambda^2\ \accentset{(2)}{P}_{\hat{a}}+ \lambda^4\ \accentset{(4)}{P}_{\hat{a}} + ... \ ,
\end{alignat}
\end{subequations}
where $N_0 ,  N_1$ and $N_2$ are even, odd and even natural numbers respectively. The expanded algebra is denoted by $\mathfrak{psu}(2,2|4)(N_0, N_1, N_2)$ and there are four different allowed truncation conditions, 
\begin{eqnarray}
\notag
&&N_{1} = N_{0} - 1 \, , \qquad \text{and} \qquad N_{2} = N_{0} - 2 \, , \\
\notag
&& N_{1}=N_{0}-1  \, , \qquad \text{and} \qquad N_{2}=N_{0}\, , \\
 && N_{1}=N_{0}+1   \, , \qquad \text{and} \qquad N_{2}=N_{0} \, , \\
  \notag
  && N_{1}=N_{0}+1  \, , \qquad \text{and} \qquad N_{2}=N_{0}+2 \, .
\end{eqnarray}
The commutation relations between the generators of the algebra $\mathfrak{psu}(2,2|4)(N_0, N_1, N_2)$ are inherited from the commutation relations of $\mathfrak{psu}(2,2|4)$ and share the same structure as in \autoref{expanded_comm_rel}. The commutation rules for the lower level algebra, i.e. $\mathfrak{psu}(2,2|4)(0,1,2)$,  which corresponds to the generalized \.In\"on\"u-Wigner contraction, are listed in\\ \autoref{subsec:StringySuperPoincareAlgebra}.

In order to derive the action describing the flat space regime studied in  \cite{Henneaux:1984mh} we consider the truncation at $(N_0, N_1, N_2) = (0, 1, 2)$, i.e.
\begin{equation}
J_{\hat{a}\hat{b}} = \accentset{(0)}{J}_{\hat{a}\hat{b}} \ , \qquad
Q_I = \lambda  \accentset{(1)}{Q}_I \ , \qquad
P_{\hat{a}} = \lambda^2 \accentset{(2)}{P}_{\hat{a}} \ . 
\end{equation}
By doing this, the $\mathfrak{psu}(2,2|4)$ superalgebra contracts to a subalgebra of the $\mathcal{N}=2, D=10$ Poincar\'e superalgebra\footnote{We specify that this is only a subalgebra because generators of the type $J_{a a'}$ are absent in $\mathfrak{psu}(2,2|4)$.
However, as it was shown in its construction in \cite{Henneaux:1984mh}, the action (\ref{flat_space_action}) has global symmetry the full $\mathcal{N}=2, D=10$ Poincar\'e superalgebra.}. 
The first non-zero action is 
\begin{equation}
\label{flat_space_action}
\accentset{(4)}{S} = - \frac{1}{2} \int_{\partial M_3} \text{d}^2 \sigma \sqrt{|h|} h^{\mu\nu} \accentset{(2)}{L}^{\hat{a}}_{\mu} \accentset{(2)}{L}^{\hat{b}}_{\nu} \eta_{\hat{a}\hat{b}} 
- i \int_{M_3} (\tau_3)_{IJ}  \accentset{(1)}{\bar{L}}^I \wedge \accentset{(2)}{L}^{\hat{a}} \Gamma_{\hat{a}} \wedge \accentset{(1)}{L}^J  \equiv S_{\text{flat space}}\ ,
\end{equation}
which is the flat space action constructed by Henneaux-Mezincescu \cite{Henneaux:1984mh}. 
So far we discussed the lowest order expansion, which is sufficient to reproduce the flat space action. However expansions at higher orders are interesting as well, and would produce different actions with different symmetries. It is the main part of this paper to show later that integrability and the symmetries of these higher order actions are under control.

\subsection{The Berenstein-Maldacena-Nastase (BMN) Case} 
\label{sec:BMN_limit_3form}

In the BMN limit, the spacetime geometry seen by the fast spinning point-like string changes from AdS$_5 \times$S$^5$ to the pp-wave background.  The action in the pp-wave background has been constructed by Metsaev \cite{Metsaev:2001bj}, and the contraction that leads from the $\mathfrak{psu}(2,2|4)$ superalgera to the pp-wave isometry superalgebra has been found by Hatsuda, Kamimura and Sakaguchi \cite{Hatsuda:2002xp}.\\ 

The reader should refer to \cite{Hatsuda:2002xp} for the notation used in this section. Here the space indices run as $\hat{i}, \hat{j}, ... = 1, ... , 8$, while $+, -$ indicate the light-cone directions. 
The generator $P^*_{\hat{i}}$ is defined as $P^*_{\hat{i}} \equiv J_{0\hat{i}}$ if $\hat{i} = 1, ..., 4$, or as $P^*_{\hat{i}} \equiv J_{9\hat{i}}$ if $\hat{i} = 5, ..., 8$, and $Q_{\pm , I}$ are the projected supercharges defined by \cite{Hatsuda:2002xp}
\begin{align}
Q_{\pm,I}=\frac{1}{2}(1\pm \Gamma_{9}\Gamma_{0})Q_{I}.
\end{align}
The commutation relations are listed in \autoref{subsec:BMNAlgebra}.
We decompose the $\mathfrak{psu}(2,2|4)$ superalgera as\footnote{The relation between the expansion parameter $\lambda$ and the contraction parameter $\Omega$ in \cite{Hatsuda:2002xp} is $\lambda = \Omega^{-1}$. } 
\begin{align}
\mathfrak{psu}(2,2|4) = V_0 \oplus V_1 \oplus V_2  \, ,
\end{align}
where
\begin{multieqB}[3]
V_0&= \text{span}\{ J_{\hat{i}\hat{j}} , P_- ,  Q_{- , I} \}\, , \nonumber\\
V_1 &= \text{span}\{ P_{\hat{i}}, P^*_{\hat{i}} , Q_{+ , I} \} \, , \nonumber\\
V_2 &= \text{span}\{ P_+ \} \, . 
\end{multieqB}
\noindent The decomposition above induces the following expansion of the generators
\begin{subequations}
\begin{alignat}{2}
J_{\hat{i}\hat{j}}=&\sum_{\substack{k=0,\ k\text{ even}}}^{N_{0}}\lambda^{k}\ \accentset{(k)}{J}_{\hat{i}\hat{j}}&&=\accentset{(0)}{J}_{\hat{i}\hat{j}}+\lambda^2\ \accentset{(2)}{J}_{\hat{i}\hat{j}}+... \ ,\\
P_{-}=&\sum_{\substack{k=0,\ k\text{ even}}}^{N_{0}}\lambda^{k}\ \accentset{(k)}{P}_{-}&&=\accentset{(0)}{P}_{-}+\lambda^2\ \accentset{(2)}{P}_{-}+... \ ,\\
Q_{-,I}=&\sum_{\substack{k=0,\ k\text{ even}}}^{N_{0}}\lambda^{k}\ \accentset{(k)}{Q}_{-,I}&&=\accentset{(0)}{Q}_{-,I}+\lambda^2\ \accentset{(2)}{Q}_{-,I}+... \ ,\\
P_{\hat{i}} =&\sum_{\substack{k=1,\ k\text{ odd}}}^{N_{1}}\lambda^{k}\ \accentset{(k)}{P}_{\hat{i}} &&=\lambda\ \accentset{(1)}{P}_{\hat{i}} +\lambda^3\ \accentset{(3)}{P}_{\hat{i}}+... \ ,\\
P^{*}_{\hat{i}} =&\sum_{\substack{k=1,\ k\text{ odd}}}^{N_{1}}\lambda^{k}\ \accentset{(k)}{P}^{*}_{\hat{i}} &&=\lambda\ \accentset{(1)}{P}^{*}_{\hat{i}} +\lambda^3\ \accentset{(3)}{P}^{*}_{\hat{i}}+... \ ,\\
Q_{+,I} =&\sum_{\substack{k=1,\ k\text{ odd}}}^{N_{1}}\lambda^{k}\ \accentset{(k)}{Q}_{+,I} &&=\lambda\ \accentset{(1)}{Q}_{+,I} +\lambda^3\ \accentset{(3)}{Q}_{+,I}+... \ ,\\
P_{+} =&\sum_{\substack{k=2,\ k\text{ even}}}^{N_{2}}\lambda^{k}\ \accentset{(k)}{P}_{+} &&=\lambda^2\ \accentset{(2)}{P}_{+} +\lambda^4\ \accentset{(4)}{P}_{+}+... \ ,
\end{alignat}
\end{subequations}
with truncation conditions
\begin{eqnarray}
\notag
&&N_{1} = N_{0} - 1 \, , \qquad \text{and} \qquad N_{2} = N_{0} \, , \\
 && N_{1}=N_{0}+1   \, , \qquad \text{and} \qquad N_{2}=N_{0} \, , \\
  \notag
  && N_{1}=N_{0}+1  \, , \qquad \text{and} \qquad N_{2}=N_{0}+2 \, .
\end{eqnarray}
By truncating at $(N_0 , N_1, N_2) = (0, 1, 2)$, 
the $\mathfrak{psu}(2,2|4)$ superalgebra contracts to the pp-wave isometry superalgebra.  The first non-zero action is 
\begin{eqnarray}
\label{pp_wave_action}
\notag
\accentset{(2)}{S} &=& - \frac{1}{2}\int_{\partial M_3} \text{d}^2 \sigma \sqrt{|h|} h^{\mu\nu} \bigg( 2 \accentset{(2)}{L}^+_{\mu} \accentset{(0)}{L}^-_{\nu} + \accentset{(1)}{L}^{\, \hat{i}}_{\mu}\accentset{(1)}{L}^{\, \hat{j}}_{\nu} \delta_{\hat{i}\hat{j}} \bigg)  - i \int_{M_3} (\tau_3)_{IJ} \bigg( \accentset{(0)}{\bar{L}}^{- , I} \wedge \accentset{(2)}{L}^+ \Gamma_+ \wedge \accentset{(0)}{L}^{- , J} \\
&+& \accentset{(1)}{\bar{L}}^{+ , I} \wedge \accentset{(0)}{L}^- \Gamma_- \wedge \accentset{(1)}{L}^{+ , J} 
+ \accentset{(1)}{\bar{L}}^{+ , I} \wedge \accentset{(1)}{L}^{\, \hat{i}} \Gamma_{\hat{i}} \wedge \accentset{(0)}{L}^{- , J} 
+ \accentset{(0)}{\bar{L}}^{- , I} \wedge \accentset{(1)}{L}^{\, \hat{i}} \Gamma_{\hat{i}} \wedge \accentset{(1)}{L}^{+ , J} \bigg) \equiv S_{\text{BMN}}\, ,
\end{eqnarray}
and by recalling the properties of the light-cone projectors, the Wess-Zumino term of \autoref{pp_wave_action} can be recombined into the form $(\tau )_{IJ} \bar{L}^I \wedge L^{\hat{a}} \Gamma_{\hat{a}} \wedge L^J$, which is in agreement with Metsaev \cite{Metsaev:2001bj}. Also in this case, one can consider higher order expansions which go beyond the contraction.

\subsection{The Non-Relativistic Cases}

\subsubsection{The Newton-Hooke Case}
\label{sec:non_rel}
A non-relativistic limit of the AdS$_5 \times$S$^5$ action has been taken in\cite{Gomis:2005pg}. In our approach we take an expansion procedure instead of a limit, and at the end of this section we comment how we reproduce the result of \cite{Gomis:2005pg}.

To deal with the non-relativistic cases which will be discussed in this paper, we decompose the AdS$_{5}$ index as $a=(\overline{a}, \underline{a})$ where $\overline{a}=0,1$ and $\underline{a}=2,3,4$. $\overline{a}$ are longitudinal indices, $\underline{a}$ transverse indices with respect to the string worldsheet.\footnote{The name longitudinal and transverse is defined in the context of non-relativistic theory by the fact that in this regime the space is foliated \cite{Gomis:2000bd, Bergshoeff:2019pij, Bergshoeff:2018yvt}.} In this section we are interested in the Newton-Hooke non-relativistic case, which is characterised by the fact that the subspace decomposition of $\mathfrak{psu}(2,2|4)$ defines an expansion such that at the lowest level one gets the \.In\"on\"u-Wigner contraction that leads from super-AdS algebra to super-Newton-Hooke algebra.
To do this, we need to project the supercharges as
\begin{align}
Q_{\pm , I} &=  (\hat{\Pi}_{\pm})_{IJ} Q_J \, ,\label{eq:Qprojection}
\end{align}
where
\begin{align}
(\hat{\Pi}_{\pm})_{IJ} &=\frac{1}{2}\left[1\pm \Gamma_{0}\Gamma_{1} \otimes (\tau_3)_{IJ} \right].
\end{align}
Next, we decompose the $\mathfrak{psu}(2,2|4)$ superalgebra as\footnote{In this case, the expansion parameter $\lambda$ cannot be identified straightforwardly with the expansion parameter $\omega$ in \cite{Gomis:2005pg}, since in the method proposed in this paper negative powers of $\lambda$ are not considered. However the actions are related by a global rescaling factor.}
\begin{align}
\mathfrak{psu}(2,2|4) = V_0 \oplus V_1 \, ,\nonumber
\end{align}
where
\begin{align}
\label{NH_decomp}
V_0 &= \text{span} \{ J_{\overline{a}\overline{b}},\ J_{\underline{a}\underline{b}},\ J_{a'b'},\  P_{\overline{a}},\ Q_{+ , I} \} \, , \nonumber\\
V_1 &= \text{span} \{J_{\overline{a}\underline{b}},\ P_{\underline{a}} ,\ P_{a'} ,\ Q_{- , I} \} \, ,
\end{align}
The expansion follows the general rules described in \autoref{Sec:expansion_Z2}. We perform the truncation at $(N_0 , N_1) = (2, 1)$, i.e.
\begin{multieqB}[3]
J_{\overline{a}\overline{b}} &= \accentset{(0)}{J}_{\overline{a}\overline{b}} + \lambda^2 \accentset{(2)}{J}_{\overline{a}\overline{b}}\, , \nonumber\\
J_{\underline{a}\underline{b}} &= \accentset{(0)}{J}_{\underline{a}\underline{b}} + \lambda^2 \accentset{(2)}{J}_{\underline{a}\underline{b}} \, , \nonumber\\
J_{a'b'} &= \accentset{(0)}{J}_{a'b'} + \lambda^2 \accentset{(2)}{J}_{a'b'}\, ,\nonumber\\
P_{\overline{a}} &= \accentset{(0)}{P}_{\overline{a}} + \lambda^2 \accentset{(2)}{P}_{\overline{a}} \, ,\nonumber\\
Q_{+ , I} &= \accentset{(0)}{Q}_{+ , I} + \lambda^2 \accentset{(2)}{Q}_{+ , I}\, ,\nonumber\\
J_{\overline{a}\underline{b}} &= \lambda\accentset{(1)}{J}_{\overline{a}\underline{b}}\, , \nonumber\\
P_{\underline{a}} &= \lambda\accentset{(1)}{P}_{\underline{a}} \, ,\nonumber\\
P_{a'} &= \lambda\accentset{(1)}{P}_{a'} \, ,\nonumber\\
Q_{- , I} &= \lambda\accentset{(1)}{Q}_{-, I}\, .\label{eq:NHexpansion}
\end{multieqB}
\noindent By expanding, the $\mathfrak{psu}(2,2|4)$ superalgebra becomes an extension of the stringy Newton-Hooke superalgebra.  The first non-zero action is 
\begin{equation}
\label{newton_hooke_zeroth_action}
\accentset{(0)}{S} = - \frac{1}{2}\int_{\partial M_3} \text{d}^2 \sigma \sqrt{|h|} h^{\mu\nu} 
\accentset{(0)}{L}^{\overline{a}}_{\mu} \accentset{(0)}{L}^{\overline{b}}_{\nu} \eta_{\overline{a}\overline{b}} 
- i \int_{M_3} (\tau_3)_{IJ} \accentset{(0)}{\bar{L}}^{+ , I} \wedge \accentset{(0)}{L}^{\overline{a}}\Gamma_{\overline{a}}\wedge \accentset{(0)}{L}^{+ , J} \equiv S^{\text{div}}_{\text{non-rel}}\, , 
\end{equation}
which reproduces the superficially divergent term of \cite{Gomis:2005pg}, while the first non-zero action associated with a non-zero power of $\lambda$ is
\begin{eqnarray}
\label{newton_hooke_second_action}
\notag
\accentset{(2)}{S} &=& - \frac{1}{2}\int_{\partial M_3} \text{d}^2 \sigma \sqrt{|h|} h^{\mu\nu} \bigg( 2 \accentset{(2)}{L}^{\overline{a}}_{\mu} \accentset{(0)}{L}^{\overline{b}}_{\nu} \eta_{\overline{a}\overline{b}} + \accentset{(1)}{L}^{\underline{a}}_{\mu} \accentset{(1)}{L}^{\underline{b}}_{\nu} \delta_{\underline{a}\underline{b}} + \accentset{(1)}{L}^{a'}_{\mu} \accentset{(1)}{L}^{b'}_{\nu} \delta_{a' b'} \bigg) \\
\notag
&-& i \int_{M_3} (\tau_3)_{IJ} \bigg( \accentset{(0)}{\bar{L}}^{+ , I} \wedge \accentset{(2)}{L}^{\overline{a}} \Gamma_{\overline{a}} \wedge \accentset{(0)}{L}^{+ , J} 
+ \accentset{(1)}{\bar{L}}^{- , I} \wedge \accentset{(0)}{L}^{\overline{a}} \Gamma_{\overline{a}
} \wedge \accentset{(1)}{L}^{- , J} + 2 \accentset{(2)}{\bar{L}}^{+ , I} \wedge \accentset{(0)}{L}^{\overline{a}} \Gamma_{\overline{a}} \wedge \accentset{(0)}{L}^{+ , J} \\
&+& 2 \accentset{(1)}{\bar{L}}^{- , I} \wedge \accentset{(1)}{L}^{\underline{a}} \Gamma_{\underline{a}} \wedge \accentset{(0)}{L}^{+ , J} 
+ 2 \accentset{(1)}{\bar{L}}^{- , I} \wedge \accentset{(1)}{L}^{a'} \Gamma_{a'} \wedge \accentset{(0)}{L}^{+ , J}  
 \bigg) \equiv S^{\text{finite}}_{\text{non-rel}} \, , 
\end{eqnarray}
which matches the finite term of \cite{Gomis:2005pg}, up to the term $\accentset{(2)}{\bar{L}}^{+ , I} \wedge \accentset{(0)}{L}^{\overline{a}} \Gamma_{\overline{a}} \wedge \accentset{(0)}{L}^{+ , J}$. 
The comparison with \cite{Gomis:2005pg} should only be made before any limit is taken (i.e. one can make a comparison with equations (4.2) to (4.5) therein, up to the power shift of $\lambda$). 
The global symmetry appearing in \cite{Gomis:2005pg} is the stringy Newton-Hooke superalgebra, while in our case the actions \autoref{newton_hooke_zeroth_action} and \autoref{newton_hooke_second_action} are invariant under the \emph{extended} stringy Newton-Hooke superalgebra, where in \autoref{newton_hooke_second_action} it acts fully, while in \autoref{newton_hooke_zeroth_action} in part trivially. 
This is a reflection of the fact that in \cite{Gomis:2005pg} the fields are rescaled by a parameter $\omega$, which in turns implies an expansion of the Maurer-Cartan 1-forms. However the expanded terms in the Maurer-Cartan equations do not gain the meaning of coming from new algebra generators, which is on the other hand the main point of the Lie algebra expansion. For this reason our model has 12 bosonic fields, while in \cite{Gomis:2005pg} there are only 10. However we expect that one can truncate our theory to a subsector, where for instance in the bosonic sector the two fields associated with $\accentset{(2)}{P}_{\overline{a}}$ are identified with the two fields associated with $\accentset{(0)}{P}_{\overline{a}}$. In this way we expect that the global symmetry can also be reduced to the stringy Newton-Hooke superalgebra.

\subsubsection{The Galilei Case}
In this section we propose the Galilean AdS$_5 \times$S$^5$ action by using the method of Lie algebra expansion. This case can be obtained in two different ways, one is to start from the Newton-Hooke case and perform a large radius expansion, the other one is to start from the large radius expansion and perform a non-relativistic expansion. 
We remark that the commutativity of the two expansions are meant at the level of the contraction, while for higher orders, some attention should be paid in comparing the two procedures.  Here we follow the pattern from super-Poincar\'e to super-Galilei. The decomposition to consider is the same leading from $\mathfrak{psu}(2,2|4)$ to super-Newton-Hooke, i.e. 
\begin{align}
\text{Super-Poincaré} = V_0 \oplus V_1 \nonumber
\end{align}
where
\begin{align}
V_0 &= \text{span} \{ J_{\overline{a}\overline{b}},\ J_{\underline{a}\underline{b}},\ J_{a'b'},\  P_{\overline{a}},\ Q_{+ , I} \} \, ,\nonumber\\
V_1 &= \text{span} \{J_{\overline{a}\underline{b}},\ P_{\underline{a}} ,\ P_{a'} ,\ Q_{- , I} \} \, .
\end{align}
The projected supercharges $Q_{\pm , I}$ are defined as in \autoref{eq:Qprojection} and the expansion follows the same rules of  \autoref{eq:NHexpansion}. Performing a truncation at $(N_{0},\ N_{1})=(2,1)$, the super-Poincar\'e algebra becomes an extension of the super-Galilei algebra. The lowest order action is
\begin{equation}
\label{galilei_zeroth_action}
\accentset{(0)}{S} = - \frac{1}{2}\int_{\partial M_3} \text{d}^2 \sigma \sqrt{|h|} h^{\mu\nu} 
\accentset{(0)}{L}^{\overline{a}}_{\mu} \accentset{(0)}{L}^{\overline{b}}_{\nu} \eta_{\overline{a}\overline{b}} 
- i \int_{M_3} (\tau_3)_{IJ} \accentset{(0)}{\bar{L}}^{+ , I} \wedge \accentset{(0)}{L}^{\overline{a}}\Gamma_{\overline{a}}\wedge \accentset{(0)}{L}^{+ , J} \, , 
\end{equation}
and at the next order, we have
\begin{eqnarray}
\label{galilei_second_action}
\notag
\accentset{(2)}{S} &=& - \frac{1}{2}\int_{\partial M_3} \text{d}^2 \sigma \sqrt{|h|} h^{\mu\nu} \bigg( 2 \accentset{(2)}{L}^{\overline{a}}_{\mu} \accentset{(0)}{L}^{\overline{b}}_{\nu} \eta_{\overline{a}\overline{b}} + \accentset{(1)}{L}^{\underline{a}}_{\mu} \accentset{(1)}{L}^{\underline{b}}_{\nu} \delta_{\underline{a}\underline{b}} + \accentset{(1)}{L}^{a'}_{\mu} \accentset{(1)}{L}^{b'}_{\nu} \delta_{a' b'} \bigg) \\
\notag
&-& i \int_{M_3} (\tau_3)_{IJ} \bigg( \accentset{(0)}{\bar{L}}^{+ , I} \wedge \accentset{(2)}{L}^{\overline{a}} \Gamma_{\overline{a}} \wedge \accentset{(0)}{L}^{+ , J} 
+ \accentset{(1)}{\bar{L}}^{- , I} \wedge \accentset{(0)}{L}^{\overline{a}} \Gamma_{\overline{a}} \wedge \accentset{(1)}{L}^{- , J} + 2 \accentset{(2)}{\bar{L}}^{+ , I} \wedge \accentset{(0)}{L}^{\overline{a}} \Gamma_{\overline{a}} \wedge \accentset{(0)}{L}^{+ , J} \\
&+& 2 \accentset{(1)}{\bar{L}}^{- , I} \wedge \accentset{(1)}{L}^{\underline{a}} \Gamma_{\underline{a}} \wedge \accentset{(0)}{L}^{+ , J} 
+ 2 \accentset{(1)}{\bar{L}}^{- , I} \wedge \accentset{(1)}{L}^{a'} \Gamma_{a'} \wedge \accentset{(0)}{L}^{+ , J}  
 \bigg)  \, , 
\end{eqnarray}
We remark that these actions look formally the same as the ones in the super-Newton-Hooke case, although the global symmetry is different. 
Higher order truncations of the non-relativistic cases here discussed could also be considered.

\section{Lie Algebra Expansion and Integrability }
\label{sec:integrability}
In this section we study the integrability properties of the $\sigma$ models associated with the expanded algebras. We define the conditions under which the final model is integrable and derive the expression for the corresponding Lax connection.\\

Consider a string $\sigma$-model action $S$ given in (\ref{action_master1}) expressed in terms of the fields, $\alpha^{a}$, where $a$ is a generic multi-index (e.g. the $\mathbb{Z}_{4}$ grading, the generator's label, or both).
Denoting the equations of motion by $\mathcal{E}_{a}$, we write the variation of the action as
\begin{align}
\delta S=&\mathcal{E}^{a}\delta \alpha_{a}\, ,
\end{align}
where the sum over $a$ is understood.  We denote the Lax connection by
\begin{align}
\mathscr{L}&= \mathpzc{k}_{\ a}\mathcal{A}^{a} + \mathpzc{k}^{'}_{\ a}\star_2\mathcal{A}^{a} \, ,
\end{align}
where $\mathpzc{k}_{\ a}, \mathpzc{k}^{'}_{\ a}$ are some coefficients, which in general are functions of a spectral parameter.
We have, by initial assumption, that the equations of motion are equivalent to the zero curvature condition for the Lax pair , i.e.
\begin{align}
\mathcal{Z}=\text{d}\mathscr{L}-\frac{1}{2}\mathscr{L}\wedge\mathscr{L}=0 \, .
\end{align}
The equivalence has to be understood between the set of equations of motion, with index $a$, and the set of zero curvature equations organized with the same index, i.e.
\begin{align}
\Big\{\mathcal{E}^{a}=0\Big\} \quad \Longleftrightarrow \quad\mathcal{Z}=0\, .
\end{align}
Now we can expand both sides and the equivalence should continue to hold, furthermore it should hold order by order, i.e.
\begin{align}
\Big\{\accentset{(k)}{\mathcal{E}}^{a}=0\Big\} \quad\Longleftrightarrow \quad\accentset{(k)}{\mathcal{Z}}=0 \, ,\label{eq:correspondence}
\end{align}
namely the set of equations of motion at order $k$ in the expansion parameter is equivalent to the set of conditions coming from the zero curvature equation at the same order.  However, the expansion is applied at the level of the action, and not at the level of the equations of motion. This means that in the variation of a certain expanded action $\accentset{(n)}{S}$ it appears both the expansion of the equations of motion and the order in $\lambda$ of the associated field. Moreover, it can be  checked explicitly that the equations of motion for the $\accentset{(n)}{S}$ action obtained from the expansion of (\ref{action_master1}) coincides with the equations of motion for (\ref{action_master1}) expanded up to the $n$-th order. Therefore we have that
\begin{align}
\delta\mathcal{S}=\sum_{k=0}^{\infty}\lambda^{k}\bigg(\sum_{i+j=k}\accentset{(i)}{\mathcal{E}}^{a}\delta\accentset{(j)}{ \alpha}_{a}\bigg)=0 \, ,
\end{align}
which must be compared with the expansion of the zero curvature equation
\begin{align}
\sum_{k=0}^{\infty}\lambda^{k}v_{a}\accentset{(k)}{\mathcal{Z}}^{a}=0 \, ,
\end{align}
where, in view of the expansion to not carry free indices, we contracted the zero curvature condition with a set of orthonormal vectors $v_{a}$, which has just the meaning that any $a$ components of $\mathcal{Z}$ should satisfy the equation separately.
In order to study the general behaviour, it is useful to write down explicitly all terms entering in the expansions above, i.e.
\begin{align}
\accentset{(0)}{\mathcal{E}}^{a}\delta\accentset{(0)}{ \alpha}_{a}+\lambda \Big(\accentset{(1)}{\mathcal{E}}^{a}\delta\accentset{(0)}{ \alpha}_{a}+\accentset{(0)}{\mathcal{E}}^{a}\delta\accentset{(1)}{ \alpha}_{a}\Big)+\lambda^{2}\Big(\accentset{(0)}{\mathcal{E}}^{a}\delta\accentset{(2)}{ \alpha}_{a}+\accentset{(1)}{\mathcal{E}}^{a}\delta\accentset{(1)}{ \alpha}_{a}+\accentset{(2)}{\mathcal{E}}^{a}\delta\accentset{(0)}{ \alpha}_{a}\Big)+...=0\, ,
\end{align}
and
\begin{align}
v_{a}\accentset{(0)}{\mathcal{Z}}^{a}+\lambda v_{a}\accentset{(1)}{\mathcal{Z}}^{a}+\lambda^{2}v_{a}\accentset{(2)}{\mathcal{Z}}^{a}+...=0 \, .
\end{align}
Comparing order by order we obtain the following table.
\begin{table}[h]
\centering
\renewcommand{\arraystretch}{2}
\begin{tabular}{crcl}
{\bf Order}&{\centering \bf Variation of the Action}&&{\bf \centering Zero Curvature Equations}\\
\hline
0&$\accentset{(0)}{\mathcal{E}}_{a}\delta\accentset{(0)}{ \alpha}^{a}=0$&$\phantom{\Longleftrightarrow}$&$v_{a}\accentset{(0)}{\mathcal{Z}}^{a}=0$\\
1&$\Big(\accentset{(1)}{\mathcal{E}}_{a}\delta\accentset{(0)}{ \alpha}^{a}+\accentset{(0)}{\mathcal{E}}_{a}\delta\accentset{(1)}{ \alpha}^{a}\Big)=0$&$\phantom{\Longleftrightarrow}$&$ v_{a}\accentset{(1)}{\mathcal{Z}}^{a}=0$\\
2&$\Big(\accentset{(0)}{\mathcal{E}}_{a}\delta\accentset{(2)}{ \alpha}^{a}+\accentset{(1)}{\mathcal{E}}_{a}\delta\accentset{(1)}{ \alpha}^{a}+\accentset{(2)}{\mathcal{E}}_{a}\delta\accentset{(0)}{ \alpha}^{a}\Big)=0$&$\phantom{\Longleftrightarrow}$&$v_{a}\accentset{(2)}{\mathcal{Z}}^{a}=0$\\
&$\vdots$&&$\vdots$\\
$n$&$\bigg(\sum_{i=0}^{n}\accentset{(i)}{\mathcal{E}}_{a}\delta\accentset{(n-i)}{ \alpha^{a}}\ \bigg)=0$&$\phantom{\Longleftrightarrow}$&$ v_{a}\accentset{(n)}{\mathcal{Z}}^{a}=0$\\ [0.3cm]
\hline
\end{tabular}
\end{table}

\FloatBarrier
By using the equivalence \autoref{eq:correspondence}, we recognise that at zeroth order the set of equations $\accentset{(0)}{\mathcal{E}}^{a}=0$ is equivalent to the set of equations $\accentset{(0)}{\mathcal{Z}}^{a}=0$. Moving at first order, we recognise that the equations of motion for $\accentset{(1)}{ \alpha}_{a}$ are equivalent to the zeroth order equations $\accentset{(0)}{\mathcal{Z}}^{a}=0$, while the equations of motion for $\accentset{(0)}{ \alpha}_{a}$ are equivalent to the zero curvature condition at first order, $\accentset{(1)}{\mathcal{Z}}^{a}=0$. If we consider now the equations of motion for the second order action then we find that they are equivalent to the zero curvature equations up to the second order. This is a general recursive relation. Thus we learn that considering the expanded action at order $n$,  $\accentset{(n)}{S}$, the equations of motion for this action are equivalent to the zero curvature condition up to that order.

This argument suggests us that the initial Lax connection expanded up to order $n$ is a good candidate Lax connection for the action $\accentset{(n)}{S}$. However in order to apply the general correspondence above to our models there are some subtleties that we have to consider. In particular it could happen that, due to the specific expansion or truncation, some fields, while appearing in the zero curvature equations,  may not appear in the action and thus we loose the corresponding equations of motion. This implies that, in these cases, we cannot consider all the zero curvature equations up to order $n$,  but we need to refine the correspondence by imposing further conditions. In order to have an equivalence between the zero curvature equations and the equations of motion we need to make sure that the equations of motion coming from the action contain the same set of Maurer Cartan form components appearing in the Lax connection. Considering a truncated algebra $\mathfrak{g}(N_{0},N_{1},...)$ the associated Lax connection is obtained by expanding the original Lax connection up to the order of truncation. This implies that all the currents appearing in the algebra will also appear in it and in the zero curvature equations, thus in order to preserve the equivalence with the equations of motion, after the expansion, we should require the same to happen in the $n$-th order term Lagrangian.\footnote{We remark that the quadratic term in zero curvature equation could not produce term of  order higher than the truncation order, since commutation relations between expanded generators whose sum of expansion order exceeds the truncation order vanish identically.} We are going to inspect how this request will put some constraints on the possible choices of $n$ in relation with the given truncation $N_{0},\ N_{1},\ ...$ .

There are two possible mechanisms that could affect the correspondence between equations of motions and zero curvature equations , the first due to a shift in the lowest order of the expansion, the second due to the truncation. In order to understand them and derive the opportune conditions let us consider a simplified model, a Lagrangian term given by two generic currents, $\mathcal{A}$ and $\mathcal{B}$, 
\begin{align}
\mathcal{L}&=\mathcal{A} \mathcal{B}\ ,
\end{align}
where contractions of the indices carried by the two currents, with opportune tensors, are understood; these will not play any role in this context. All the cases we are dealing with in the present work fall in the following type of expansion
\begin{subequations}
\begin{align}
\mathcal{A}&=\lambda^{n_{A}}\accentset{(n_{A})}{\mathcal{A}}+\lambda^{n_{A}+1}\ \accentset{(n_{A}+1)}{\mathcal{A}}+...\\
\mathcal{B}&=\lambda^{n_{B}}\accentset{(n_{B})}{\mathcal{B}}+\lambda^{n_{B}+1}\ \accentset{(n_{B}+1)}{\mathcal{B}}+... \ ,
\end{align}\label{eq:expansionAB}
\end{subequations}
where $n_{A}, n_{B}$ are the lowest order expansion for currents $\mathcal{A}$ and $\mathcal{B}$ respectively and are non-negative integers. We can plug the expansion in the Lagrangian obtaining 
\begin{align}
\mathcal{L}&=\lambda^{n_{A}+n_{B}}\accentset{(n_{A})}{\mathcal{A}}\quad \accentset{(n_{B})}{\mathcal{B}}+\lambda^{n_{A}+n_{B}+1}\Big(\quad \accentset{(n_{A}+1)}{\mathcal{A}}\quad\accentset{(n_{B})}{\mathcal{B}}+\accentset{(n_{A})}{\mathcal{A}}\quad \accentset{(n_{B}+1)}{\mathcal{B}}\quad \Big)+...\ .
\end{align}
The first mechanism spoiling the equivalence could be already understood looking at the lowest order term, at order $n_{A}+n_{B}$. This contains only the currents $\accentset{(n_{A})}{\mathcal{A}}$ and $\accentset{(n_{B})}{\mathcal{B}}$, thus if we have considered the Lax connection up to this order, as above, we would have naively included in it all the currents up to the order $n_{A}+n_{B}$, but this means that if $n_{B}\geqslant 1$ we would have considered also $\accentset{(n_{A}+1)}{\mathcal{A}}$. This current clearly cannot be present in the Lax connection associated with the lowest order term in the action since this does not produce the corresponding equation of motion. Thus we have a shift between the order of the Lagrangian term we are considering and the expansion of the zero curvature equations. Our strategy to take care of this problem is, rather than speaking about zero curvature equation at order $n$, to speak about zero curvature equations associated with the truncated algebra $\mathfrak{g}(N_{0},N_{1},...)$ and then require that all the fields of the algebra appear in the $n$-th order Lagrangian. Thus instead of starting from the action and then build the Lax connection we start from the Lax connection associated with a truncated algebra $\mathfrak{g}(N_{0},\ N_{1},\ ...)$, i.e. obtained by expanding the original Lax connection $\mathscr{L}$ up to the truncation orders, denoted with 
\begin{align}
\mathscr{L}(N_{0},\ N_{1},\ ... )\ ,
\end{align}
and then ask ourselves if there is an action term in the expansion of the Lagrangian whose equations of motion are equivalent to the zero curvature equation of $\mathscr{L}(N_{0},\ N_{1},\ ... )$. In particular this means to investigate if there is an action term at a certain order $n$ containing all the currents appearing in the Lax connection $\mathscr{L}(N_{0},\ N_{1},\ ... )$. To do this we could consider the $n=m+n_{A}+n_{B}$ order term, where $m$ is a generic natural number, in the Lagrangian above, 
\begin{align}
\accentset{(m+n_{A}+n_{B})}{\mathcal{L}}&\quad =\quad \accentset{(m+n_{A})}{\mathcal{A}}\ \quad \accentset{(n_{B})}{\mathcal{B}}\ +\ \accentset{(m+n_{A}-1)}{\mathcal{A}}\qquad \accentset{(n_{B}+1)}{\mathcal{B}}\ +...+\ \accentset{(n_{A})}{\mathcal{A}}\ \quad \accentset{(n_{B}+m)}{\mathcal{B}}\ .\label{eq:termnAB}
\end{align} 
This term contains the $\mathcal{A}$ and $\mathcal{B}$ currents at order $i_{A}$ and $i_{B}$ respectively, with 
\begin{subequations}
\begin{align}
n_{A}&\leqslant i_{A}\leqslant n-n_{B}\\
n_{B}&\leqslant i_{B}\leqslant n-n_{A}\ .
\end{align}
\end{subequations} 
The second mechanism that could affect our correspondence is due to the truncation of the infinite expansion.  Truncating the expansion of $\mathcal{A}$ and $\mathcal{B}$ at orders $N_{A}$ and $N_{B}$ respectively then if $N_{A}> n-n_{B}$ the current $\mathcal{A}$ at order $N_{A}$ cannot appear \autoref{eq:termnAB}. The same is true for the current $\mathcal{B}$ at order $N_{B}$ if $N_{B}> n-n_{A}$ and depending on the specific truncation more terms could not be present in the Lagrangian term while being part of the truncated algebra. In particular the set of terms appearing in \autoref{eq:termnAB} after the truncation is defined by the following intervals 
\begin{subequations}
\begin{align}
\max\{n_{A}, n-N_{B}\}&\leqslant i_{A}\leqslant \min\{n-n_{B}, N_{A}\}\\
\max\{n_{B}, n-N_{A}\}&\leqslant i_{B}\leqslant \min\{n-n_{A}, N_{B}\}\ .
\end{align}\label{eq:integrabilitybound}
\end{subequations} 
A missing term in the Lagrangian clearly gives a problem in the correspondence between equations of motion and zero curvature equations since, while not present in the former, it will continue to appear in the latter.  Thus we have to impose that all the currents in the algebra are present in the $n$-th order Lagrangian term after the truncation. This amount to require 
\begin{align*}
\max\{n_{A}, n-N_{B}\}&=n_{A}\\
\min\{n-n_{B}, N_{A}\}&=N_{A}\\
\max\{n_{B}, n-N_{A}\}&=n_{B}\\
\min\{n-n_{A}, N_{B}\}&=N_{B}\ ,
\end{align*}
corresponding to following conditions 
\begin{subequations}
\begin{align}
n_{A}+N_{B}&\leqslant n\leqslant N_{A}+n_{B}\\
n_{B}+N_{A}&\leqslant n\leqslant N_{B}+n_{A} \, ,
\end{align}\label{eq:conditonsimult}
\end{subequations}
were the first condition comes from the $\mathcal{A}$ current terms and the last from the $\mathcal{B}$ current terms. In particular the upper bounds of the conditions \autoref{eq:conditonsimult} are equivalent to the condition that the $n$-th order in the expansion of the Lagrangian is not modified by the truncation. The set of conditions \autoref{eq:conditonsimult} could also be expressed as
\begin{align}
\max\{n_{A}+N_{B},\ n_{B}+N_{A}\}&\leqslant n\leqslant\min\{ N_{A}+n_{B},\ N_{B}+n_{A}\}\ ,\label{eq:conditionmaxmin}
\end{align}
which in turns is equivalent to
\begin{align}
n=n_{A}+N_{B}=n_{B}+N_{A}\ .
\end{align}
We note that these conditions not only could limit the possible value of $n$ but also the possible truncations of the algebra that we could consider. The form of these conditions depends on the type of Lagrangian and expansion that we are considering, thus they should be evaluated separately in all the models we are studying. Let us just remark that the same argument could be also applied to action terms, like the Wess-Zumino term, containing more that two currents just grouping them, in turn, into two sets and repeating the analysis on all the possible sets. The final condition will be given by the intersection of all the conditions obtained in this way. 

By applying the argument just exposed to the specific cases we have considered in the present work we find that the conditions, analogous to  \autoref{eq:conditionmaxmin}, that must hold between the $n$-th order term and the truncation orders $N_{0},\ N_{1},\ ...$ to guarantee integrability for the Flat space, BMN and non-relativistic regimes are the following
\begin{description}
\item[Flat Space]
\begin{align}
\max\{N_{2}+2,\ N_{1}+3\}\leqslant &n\leqslant N_{2}+2 \label{eq:intconditionsFlat}
\end{align}
\item[BMN]
\begin{align}
\max\{N_{0}+2,\ N_{1}+1,\ N_{2}\}\leqslant &n\leqslant\min\{N_{0}+2,\ N_{1}+1,\ N_{2}\}\label{eq:intconditionsBMN}
\end{align}
\item[Newton-Hooke/Galilei]
\begin{align}
\max\{N_{0},\ N_{1}+1\}\leqslant &n\leqslant N_{0}\ . \label{eq:intconditionsNon-Relativistic}
\end{align}
\end{description}
These conditions ensure that the equations of motion coming from the $n$-th order term of the expanded action are equivalent to the zero curvature equations of the Lax connection obtained by expanding the initial Lax connection up to the truncation orders, 
\begin{align}
\mathscr{L}(N_{0},\ N_{1},\ ...)\ .
\end{align}
The results of the analysis for the different regimes we are investigating in the present work are summarized in in \autoref{tab:intcondition1} and \autoref{tab:int_k_cond}. In \autoref{tab:intcondition1},  for a given $n$-th order action term, we show the conditions that should be satisfied to preserve integrability and the truncated algebras satisfying these conditions. These algebras are the global symmetries of the $n$-th order Lagrangian term.  In \autoref{tab:int_k_cond} we show how the truncations are affected by the request of integrability of the $n$-th order action term. In particular for any possible truncation we write if this could give an integrable model and, eventually, which is the condition that the request of integrability poses between the order $n$-th and the truncation order $N_{0}$.
\begin{table}[h]
\renewcommand{\arraystretch}{3.5}
 \resizebox{\textwidth}{!}{
\begin{tabular}{cccc}
{\bf  Case}&{\centering \bf Integrability Condition }&\bf Algebra\\
\hline
 Flat &$\max\{N_{2}+2,\ N_{1}+3\}\leqslant n\leqslant N_{2}+2$&
\renewcommand{\arraystretch}{1.2}
\begin{tabular}{c}
$\mathfrak{psu}(2,2|4)(n-2,n-3,n-2)$\\
$\mathfrak{psu}(2,2|4)(n-4,n-3,n-2)$
\end{tabular}
\\
 BMN&$\max\{N_{0}+2,\ N_{1}+1,\ N_{2}\}\leqslant n\leqslant\min\{N_{0}+2,\ N_{1}+1,\ N_{2}\}$&$\mathfrak{psu}(2,2|4)(n-2,n-1,n)$\\
\renewcommand{\arraystretch}{1.2}
\begin{tabular}{c}
 Newton-Hooke/\\
 Galilei
\end{tabular}& $\max\{N_{0},\ N_{1}+1\}\leqslant n\leqslant N_{0}$&$\mathfrak{psu}(2,2|4)(n,n-1)$\\
\hline
\end{tabular}
}
\caption{\footnotesize We show, in the different cases considered in this work, for an action term at $n$-th order in the expansion, the conditions that should hold between $n$ and the truncation orders $N_{0},\ N_{1},\ ...$ to ensure integrability. In the last column we list the possible truncated algebras fulfilling these conditions. It is immediate to see that some truncations are ruled out by the conditions. }\label{tab:intcondition1}
\end{table}
\begin{table}[h]
\centering
\renewcommand{\arraystretch}{2}
\begin{tabular}{clcc}
{\bf  Case}&{\centering \bf Truncation}&&{\bf \centering Integrability Condition}\\
\hline
&$N_1 = N_0 -1$ and $N_2 = N_0 -2$&& \it not integrable\\
 &$N_1 = N_0 -1$ and $N_2 = N_0$&& $n=N_0 +2$\\
  &$N_1 = N_0 +1$ and $N_2 = N_0$&&  \it not integrable\\
\multirow{-4}{*}{ Flat Space}  &$N_1 = N_0 +1$ and $N_2 = N_0 +2$&& $n=N_0 +4$\\
  \hline
  &$N_1 = N_0 -1$ and $N_2 = N_0$&& \it not integrable\\
   &$N_1 = N_0 +1$ and $N_2 = N_0$&& \it not integrable\\
    \multirow{-3}{*}{ BMN} &$N_1 = N_0 +1$ and $N_2 = N_0 +2$&& $n = N_0+2$\\
     \hline
    &  $N_1 = N_0 -1$ && $n=N_0$ \\
 \multirow{-2}{*}{Newton-Hooke/Galilei}  &   $N_1 = N_0 +1$ &&\it not integrable \\
   \hline
\end{tabular}
\caption{\footnotesize Relations between $n$ and the truncation to be satisfied in order to have an integrable $n$-th order action. We remark that some truncation are ruled out by the request of integrabilty.}
\label{tab:int_k_cond}
\end{table}

In building an integrable expanded action and its associated Lax connection there are two possible approaches. One could start with the algebra $\mathfrak{g}(N_{0},N_{1},...)$, produced via Lie algebra expansion, then the Lax connection is fully defined, and one may wonder if there is an expanded action term satisfying the integrability relation. On the other hand one could also start from an $n$-th order action term picking it from the infinite expansion of the algebra, since the integrability condition implies that this action term will not be modified by the truncation, and then look for the truncated algebra $\mathfrak{g}(N_{0},N_{1},...)$ fulfilling the integrability conditions.  We note also that, considering an $n$-th order Lagrangian term, the algebras fulfilling the integrability condition have a fully non-trival action on it.

Having discussed the integrability of the generic expanded action, in the next subsections we give some explicit examples of Lax connection for expanded action, listing them in the different regimes studied in the previous section.
\FloatBarrier

\subsection{Integrability of Truncated Actions}

\subsubsection{The Flat Space Case}
The action $\accentset{(4)}{S}$ in \autoref{flat_space_action} is obtained from the algebra truncation at $(N_0, N_1, N_2) = (0, 1, 2)$, and therefore the condition \autoref{eq:intconditionsFlat} required for integrability is indeed satisfied. 
In the following notation for the Maurer-Cartan 1-forms
\begin{equation}
\mathcal{P}_{\mu} = L^{\hat{a}}_{\mu} P_{\hat{a}} \, , \qquad
\mathcal{Q}^I_{\mu} = L^I_{\mu} Q_{I} \, , \qquad
\mathcal{J}_{\mu} = L^{\hat{a}\hat{b}}_{\mu} J_{\hat{a}\hat{b}} \, , 
\end{equation} 
where the index $I$ is not summed over, the Lax pair for the action $\accentset{(4)}{S}$ in \autoref{flat_space_action} is 
\begin{equation}
\mathscr{L}_{\mu} = \accentset{(0)}{\mathscr{L}}_{\mu} + \accentset{(1)}{\mathscr{L}}_{\mu} + \accentset{(2)}{\mathscr{L}}_{\mu} \, , 
\end{equation}
where
\begin{eqnarray}
\notag
\accentset{(0)}{\mathscr{L}}_{\mu} &=& \ell_0 \accentset{(0)}{\mathcal{J}}_{\mu} \, , \\
\accentset{(1)}{\mathscr{L}}_{\mu} &=& \ell_3 \accentset{(1)}{\mathcal{Q}}^1_{\mu} + \ell_4  \accentset{(1)}{\mathcal{Q}}^2_{\mu} \, , \\
\notag
\accentset{(2)}{\mathscr{L}}_{\mu} &=& \ell_1 \accentset{(2)}{\mathcal{P}}_{\mu} + \ell_2 \frac{1}{\sqrt{|h|}} h_{\mu\nu} \epsilon^{\nu\rho} \accentset{(2)}{\mathcal{P}}_{\rho} \, .  
\end{eqnarray}

\subsubsection{The BMN Case}
With the truncation at $(N_0 , N_1, N_2) = (0, 1, 2)$, the action $\accentset{(2)}{S}$ in \autoref{pp_wave_action} is integrable, since \autoref{eq:intconditionsBMN} is satisfied. 
In the notation 
\begin{equation}
\mathcal{P}^{\pm}_{\mu} = L^{\pm}_{\mu} P_{\pm} \, , \qquad
\tilde{\mathcal{P}}_{\mu} = L^{\hat{i}}_{\mu} P_{\hat{i}} \, , \qquad
\mathcal{Q}^{\pm, I}_{\mu} = L^{\pm, I}_{\mu} Q_{\pm, I} \, , \qquad
\tilde{\mathcal{J}}_{\mu} = L^{\hat{i}\hat{j}}_{\mu} J_{\hat{i}\hat{j}} \, , \qquad 
\mathcal{G}^*_{\mu} = L^{*\, \hat{i}}_{\mu} P^*_{\hat{i}} \, ,
\end{equation} 
the Lax pair for the action $\accentset{(2)}{S}$ in \autoref{pp_wave_action} is 
\begin{equation}
\mathscr{L}_{\mu} = \accentset{(0)}{\mathscr{L}}_{\mu} + \accentset{(1)}{\mathscr{L}}_{\mu} +  \accentset{(2)}{\mathscr{L}}_{\mu} \, , 
\end{equation}
where
\begin{eqnarray}
\notag
\accentset{(0)}{\mathscr{L}}_{\mu} &=& \ell_0 \accentset{(0)}{\tilde{\mathcal{J}}}_{\mu} + \ell_1 \accentset{(0)}{\mathcal{P}}^-_{\mu} + \ell_2 \frac{1}{\sqrt{|h|}} h_{\mu\nu} \epsilon^{\nu\rho}\accentset{(0)}{\mathcal{P}}^-_{\rho} + \ell_3  \accentset{(0)}{\mathcal{Q}}^{-,1}_{\mu}  + \ell_4  \accentset{(0)}{\mathcal{Q}}^{-,2}_{\mu}
\, , \\
\accentset{(1)}{\mathscr{L}}_{\mu} &=& \ell_0  \accentset{(1)}{\mathcal{G}}^*_{\mu}
+ \ell_1 \accentset{(1)}{\tilde{\mathcal{P}}}_{\mu}
+ \ell_2 \frac{1}{\sqrt{|h|}} h_{\mu\nu} \epsilon^{\nu\rho}\accentset{(1)}{\tilde{\mathcal{P}}}_{\rho}
+ \ell_3  \accentset{(1)}{\mathcal{Q}}^{+,1}_{\mu}+ \ell_4 \accentset{(1)}{\mathcal{Q}}^{+,2}_{\mu} \, , \\
\notag
\accentset{(2)}{\mathscr{L}}_{\mu} &=& \ell_1 \accentset{(2)}{\mathcal{P}}^+_{\mu} + \ell_2 \frac{1}{\sqrt{|h|}} h_{\mu\nu} \epsilon^{\nu\rho} \accentset{(2)}{\mathcal{P}}^+_{\rho} \, .  
\end{eqnarray}

\subsubsection{The Newton-Hooke and Galilei Cases}
The action $\accentset{(0)}{S}$ in \autoref{newton_hooke_zeroth_action} is integrable since the condition \autoref{eq:intconditionsNon-Relativistic} is satisfied by the algebra $\mathfrak{psu}(2,2|4)(0,-1)$; this is just the subalgebra 
\begin{align}
V_0 &= \text{span} \{ J_{\overline{a}\overline{b}},\ J_{\underline{a}\underline{b}},\ J_{a'b'},\  P_{\overline{a}},\ Q_{+ , I} \} \, . 
\end{align}
By introducing the notation
\begin{eqnarray}
\notag
&&\mathcal{P}^{\parallel}_{\mu} = L^{\overline{a}}_{\mu} P_{\overline{a}} \, , \qquad
\mathcal{P}^{\perp}_{\mu} = L^{\underline{a}}_{\mu} P_{\underline{a}} \, , \qquad
\mathcal{P}^{'}_{\mu} = L^{a'}_{\mu} P_{a'} \, , \qquad 
\mathcal{Q}^{\pm, I}_{\mu} = L^{\pm, I}_{\mu} Q_{\pm, I} \, , \qquad \\ 
&&\mathcal{J}^{\parallel}_{\mu} = L^{\overline{a}\overline{b}}_{\mu} J_{\overline{a}\overline{b}} \, , \qquad 
\mathcal{J}^{\perp}_{\mu} = L^{\underline{a}\underline{b}}_{\mu} J_{\underline{a}\underline{b}} \, , \qquad
\mathcal{J}^{'}_{\mu} = L^{a'b'}_{\mu} J_{a'b'}\,\qquad 
\mathcal{G}_{\mu} = L^{\overline{a}\underline{b}}_{\mu} J_{\overline{a}\underline{b}} \, ,
\end{eqnarray}
the Lax pair for the action $\accentset{(0)}{S}$ is 
\begin{eqnarray}
\label{Lax_NH_zeroth}
\notag
\mathscr{L}_{\mu} = \accentset{(0)}{\mathscr{L}}_{\mu} &=& 
 \ell_0 (\accentset{(0)}{\mathcal{J}}^{\parallel}_{\mu} + \accentset{(0)}{\mathcal{J}}^{\perp}_{\mu} + \accentset{(0)}{\mathcal{J}}^{\, '}_{\mu} ) + \ell_1 \accentset{(0)}{\mathcal{P}}^{\parallel}_{\mu} + \ell_2 \frac{1}{\sqrt{|h|}} h_{\mu\nu} \epsilon^{\nu\rho}\accentset{(0)}{\mathcal{P}}^{\parallel}_{\rho} + \ell_3 \accentset{(0)}{\mathcal{Q}}^{+,1}_{\mu}  + \ell_4 \accentset{(0)}{\mathcal{Q}}^{+,2}_{\mu}\, . 
\end{eqnarray}
With the truncation at $(N_0 , N_1) = (2, 1)$, the action $\accentset{(2)}{S}$ in \autoref{newton_hooke_second_action} satisfies the integrability condition in \autoref{eq:intconditionsNon-Relativistic}, and its Lax pair is 
\begin{equation}
\mathscr{L}_{\mu} = \accentset{(0)}{\mathscr{L}}_{\mu} + \accentset{(1)}{\mathscr{L}}_{\mu} + \accentset{(2)}{\mathscr{L}}_{\mu} \, , 
\end{equation}
where $\accentset{(0)}{\mathscr{L}}_{\mu}$ is the same as in \autoref{Lax_NH_zeroth}, and 
\begin{eqnarray}
\notag
\accentset{(1)}{\mathscr{L}}_{\mu} &=& \ell_0  \accentset{(1)}{\mathcal{G}}_{\mu}
+ \ell_1 ( \accentset{(1)}{\mathcal{P}}^{\perp}_{\mu} + \accentset{(1)}{\mathcal{P}}^{\, '}_{\mu})
+ \ell_2 \frac{1}{\sqrt{|h|}} h_{\mu\nu} \epsilon^{\nu\rho}(\accentset{(1)}{\mathcal{P}}^{\perp}_{\rho} + \accentset{(1)}{\mathcal{P}}^{\, '}_{\rho}) +\ell_3  \accentset{(1)}{\mathcal{Q}}^{-,1}_{\mu} + \ell_4 \accentset{(1)}{\mathcal{Q}}^{-,2}_{\mu} \, , \\
\notag
\accentset{(2)}{\mathscr{L}}_{\mu} &=& \ell_0 (\accentset{(2)}{\mathcal{J}}^{\parallel}_{\mu} + \accentset{(2)}{\mathcal{J}}^{\perp}_{\mu} + \accentset{(2)}{\mathcal{J}}^{\, '}_{\mu} )
+ \ell_1 \accentset{(2)}{\mathcal{P}}^{\parallel}_{\mu} + \ell_2 \frac{1}{\sqrt{|h|}} h_{\mu\nu} \epsilon^{\nu\rho}\accentset{(2)}{\mathcal{P}}^{\parallel}_{\rho} + \ell_3 \accentset{(2)}{\mathcal{Q}}^{+,1}_{\mu}+  \ell_4  \accentset{(2)}{\mathcal{Q}}^{+,2}_{\mu} \, .
\end{eqnarray}
The same formulas presented above in the context of the Newton-Hooke case also applies for the Galilei case.

\section{Conclusions}

In this paper, the Lie algebra expansion has been applied to obtain new $\sigma$-models from a given one. The starting point is a generic 2d integrable string $\sigma$-model with a coset target space $G/H$. The isometry algebra $\mathfrak{g}$ has been expanded by using the method of the Lie algebra expansion. This in turns implies an expansion of the Maurer-Cartan 1-forms and therefore an expansion of the action. The expanded Lie algebra contains in general a greater number of generators if compared with the initial one. In our approach we associate with each generator of the new Lie algebra an independent field. This implies that the new $\sigma$-model will in general have a greater number of fields.

In the context of the AdS$_5 \times$S$^5$ $\sigma$-model, we reproduced and extended in a systematic way the action and the symmetries of some known regimes. In the flat space and BMN cases, the actions known in the literature are obtained by considering contractions of $\mathfrak{psu}(2,2|4)$. This is reproduced as a zero level expansion in our formalism, but the advantage of using the Lie algebra expansion method is that one has a complete control on symmetries and truncations also for higher order. The non-relativistic regimes of the AdS$_5 \times$S$^5$ superstring have also been explored. We reproduced the action terms in equations (4.2) to (4.5) of \cite{Gomis:2005pg}, although in our case the global symmetry is an extension of the stringy Newton-Hooke superalgebra, instead of the stringy Newton-Hooke superalgebra of \cite{Gomis:2005pg}. We commented on a possible truncation to a subsector of our model, which potentially connects our result to \cite{Gomis:2005pg}. 

The main result of this paper is about the classical integrability of the new $\sigma$-model. We gave a criterion for the algebra truncation such that the equations of motion of the actions for the new $\sigma$-model are equivalent to the vanishing curvature condition of a Lax connection. 
The strategy that we followed was to insert a given truncated expansion of the generators into the Lax pair of the initial $\sigma$-model, and to look for which $n$-th order expanded actions possess the equations of motion equivalent to the vanishing curvature of the Lax pair so constructed. This brings a set of conditions between the truncation orders $N_0, N_1, ...$ and the order $n$ of the expanded integrable action.

In the context of expansions of Chern-Simons \cite{deAzcarraga:2002xi, deAzcarraga:2019mdn} or Einstein-Hilbert \cite{Bergshoeff:2019ctr} actions, the no-missing terms relation guarantees that any gauge symmetry of the initial action is preserved also for the expanded action. The 2d $\sigma$-model action considered in this paper possesses a fermionic local symmetry, the so-called $\kappa$-symmetry, however this type of argument cannot be applied. The main reason is because the world-sheet metric $h_{\mu\nu}$ does not expand, but its $k$-symmetry variation does, since it depends on a Maurer-Cartan 1-form and a fermionic current $\kappa_{\mu}$. The $\kappa$-symmetry invariance condition of the initial action will in general expand in powers of $\lambda$, however, in contrast to what happens in \cite{deAzcarraga:2002xi, deAzcarraga:2019mdn, Bergshoeff:2019ctr}, it is now not clear how to identify it as the $\kappa$-symmetry invariance of the $n$-order action for the reason mentioned before.

We hope that this work can provoke new ideas for exploring other regimes of string $\sigma$-models. For instance, studying the expansion around the contraction that leads from the AdS superalgebra to the Caroll superalgebra is a way to investigate the ultra-relativistic limit of the string $\sigma$-model. 
Beyond classical integrability, it would be interesting to explore the \emph{quantum} integrability property of these models. By this, we mean to find the R-matrix invariant under e.g. these non- or ultra-relativistic algebras, such that it satisfies the Yang-Baxter equation, braiding unitarity and crossing-symmetry, and ultimately, writing down the Bethe ansatz by using the R-matrix so constructed\footnote{This is a program that has not been done in AdS$_5 \times$S$^5$ yet, since only the S-matrix invariant under the residual centrally extended algebra $\mathfrak{su}(2|2)^2$ has been found \cite{Beisert:2005tm}.  However this has been done for lower dimensional integrable AdS $\sigma$-models, e.g. \cite{Hoare, Fontanella:2017rvu, Sax:2012jv, Borsato:2016xns}}.

In this paper, we always expanded the string action written in terms of the Wess-Zumino (WZ) 3-form, but we did not expand the equivalent one written in terms of the WZ 2-form. The reason for this is because the equations of motion for the generic expanded action $\accentset{(n)}{S}$, obtained from an initial action $S$ written in terms of the WZ 2-form, do not coincide with the initial equations of motion for $S$ expanded up to the $n$-th order. On the other hand, this happens to be the case when the initial action $S$ is written in terms of the WZ 3-form, and this is a necessary condition in order to apply the techniques presented in this paper.

This issue has already been pointed out in \cite{Hatsuda:2001pp}, where they discuss that the algebra contraction that leads from AdS$_5 \times$S$^5$ to the flat space makes the WZ 2-form disappearing from the action, and this clearly affects the equations of motion. The authors propose that instead of taking the usual algebra contraction that leads from $\mathfrak{psu}(2,2|4)$ to the Poincar\'e superalgebra, one should take a \emph{generalised} contraction, which was the first idea of a Lie algebra expansion. In this way, one introduces an extra fermionic generator cubic in $\lambda$ which compensates the miss-matching order between the kinetic and the WZ 2-form previously found when only contracting. 
Perhaps a generalisation of this trick allows to apply the  results of this paper also to a generic action written in terms of a WZ 2-form.

Finally, it is an interesting question whether the expanded $\sigma$-model is still describing a string theory. The conditions that need to be checked are given in \cite{Zarembo:2010sg}. If the isometry algebra of the initial $\sigma$-model has vanishing Killing form, then this will persist after the expansion. Regarding the central charge condition, this should be evaluated case by case.

\vskip 0.5cm
\noindent{\bf Acknowledgements} \vskip 0.1cm
\noindent It is a pleasure to thank E. Bergshoeff,  B. Stefa\'nski, A. Torrielli, J. M. Izquierdo, F. Riccioni and especially S. van Tongeren for their feedback on the draft and for useful discussions at different stages of the work.
We also thank T. Ort\'in and D. Sorokin for interesting discussions and P. Concha, J. Hartong and N. Obers for pointing out some references. 
AF has been partially supported by the German Research Foundation (DFG) via the Emmy Noether program ``Exact Results in Extended Holography'' and by the Angelo Della Riccia Foundation Fellowship. 
LR has been supported in part by the MINECO/FEDER, UE grant PGC2018-095205-B-I00 and by the Spanish Research Agency (Agencia Estatal de Investigación) through the grant IFT Centro de Excelencia Severo Ochoa SEV-2016-0597. 
AF thanks the Instituto de F\'isica Te\'orica UAM/CSIC of Madrid for hosting his Della Riccia Fellowship.


\setcounter{section}{0}
\setcounter{subsection}{0}

\begin{appendices}

\section{Conventions}
\label{appx:convention}
We define products between Lie algebra valued 1-forms, $\mathcal{A}=A^A T_{A}$ and $\mathcal{B}=B^{A} T_{A}$ as
\begin{subequations}
\begin{align}
\mathcal{A}\wedge \mathcal{B}&= (A^A\wedge B^B)\ [T_A , T_B]\label{wedge_prod}\\
\mathcal{A}\circ \mathcal{B}&= (A^A\wedge B^B)\ T_{A} T_{B}
\end{align}\label{eq:wedgecirc}
\end{subequations}
where $A^A\wedge B^B = A^A_{\mu} B^B_{\nu} \text{d}x^{\mu} \wedge \text{d} x^{\nu}$ is the usual wedge product of differential forms. 
\vspace{1mm}
\noindent The Pauli matrices are:
\begin{equation}
\tau_1 =\left(\begin{array}{cc}
0 & 1 \\
1 & 0 
\end{array}\right) \, , \qquad
\tau_2 = \left(\begin{array}{cc}
0 & -i \\
i & 0 
\end{array}\right) \, , \qquad
\tau_3 = \left(\begin{array}{cc}
1 & 0 \\
0 & -1 
\end{array}\right) \, . 
\end{equation}
Levi-Civita symbols: 
\begin{eqnarray}
\notag
\epsilon_{\mu\nu} &:& \qquad\qquad
\epsilon^{\tau\sigma} = - \epsilon_{\tau \sigma} = + 1 \, , \\
\epsilon_{IJ} &:& \qquad\qquad
\epsilon^{12} = \epsilon_{12} = + 1 \, ,
\end{eqnarray}
Gamma matrices $\gamma_{a}$ and $\gamma_{a'}$ (associated with AdS$_5$ and S$^5$ respectively):
\begin{equation}
\{ \gamma_{a}, \gamma_{b} \} = \eta_{ab} = \text{diag}(-++++) \, , \qquad
\{ \gamma_{a'}, \gamma_{b'} \} = \delta_{a'b'} = \text{diag}(+++++) \, .
\end{equation}
Spacetime gamma matrices $\Gamma_{\hat{a}}$ :
\begin{equation}
\Gamma_{a} = \gamma_a \otimes \mathds{1}'_4 \otimes \tau_1 \, , \qquad
\Gamma_{a'} = \mathds{1}_4 \otimes  \gamma_{a'} \otimes \tau_2 \, . 
\end{equation}
The gamma matrices satisfy 
\begin{equation}
C^{-1} \Gamma_{\hat{a}} C = \Gamma_{\hat{a}}^T \, , 
\end{equation}
where $C$ is the charge conjugation matrix, which is taken to be $C= i \Gamma_0$. This implies that
\begin{equation}
\Gamma_0^T = + \Gamma_0 \, , \qquad\qquad
\Gamma_m^T = - \Gamma_m \, . 
\end{equation} 
The Majorana supercharges $Q_I$ satisfy the Weyl condition
\begin{equation}
Q_I = Q_I \frac{1 + \Gamma_{11}}{2} \, , \qquad\qquad
\Gamma_{11} = \Gamma_{0123456789} = \mathds{1}_4 \otimes \mathds{1}'_4 \otimes \tau_3 \, ,
\end{equation}

\section{Algebras}
\label{appx:algebras}
In this section we summarise the lowest order expansions which lead to the algebras that appeared through the paper. For the pp-wave algebra, we refer to \cite{Hatsuda:2002xp}.

\subsection{Super AdS$_5\times $S$^5$ Algebra: $\mathfrak{psu}(2,2|4)$}
The starting point for all expansions considered in this paper is the $\mathfrak{psu}(2,2|4)$ superalgebra, whose non zero commutation relations are
\begin{multicols}{2}
\setlength{\abovedisplayskip}{-15pt}
\allowdisplaybreaks
\begin{align}
[P_{a} , P_{b}] &= J_{ab} \nonumber\\
[P_{a}, J_{bc}] &= 2\eta_{a[b} P_{c]} \nonumber\\
[Q_I , P_{a}] &= - \frac{i}{2} \epsilon_{IJ} Q_J \gamma_{a}  \nonumber\\
[Q_{I},J_{ab}]&=-\frac{1}{2}Q_{I}\gamma_{ab} \nonumber\\
[P_{a'} , P_{b'}] &= - J_{a'b'}  \nonumber\\
[P_{a'}, J_{b'c'}] &= 2\eta_{a'[b'} P_{c']}  \nonumber\\
[Q_I , P_{a'}] &=  \frac{1}{2} \epsilon_{IJ} Q_J \gamma_{a'} \nonumber\\
[Q_{I},J_{a'b'}]&=-\frac{1}{2}Q_{I}\gamma_{a'b'} \nonumber
\end{align}
\end{multicols}
\setlength{\parindent}{0pt}
\vspace{-0.9cm}
\begin{align}
[J_{ab} , J_{cd}] &= 2\eta_{c[b}J_{a]d}-2\eta_{d[b}J_{a]c} \nonumber\\ 
[J_{a'b'} , J_{c'd'}] &= 2\eta_{c'[b'}J_{a']d'} -2\eta_{d'[b'}J_{a']c'} \nonumber \\
\{ Q_{\alpha\alpha'I} , Q_{\beta\beta'J} \} &= \delta_{IJ} \bigg[ - 2i C_{\alpha'\beta'} (C \gamma^{a} )_{\alpha\beta} P_{a} + 2 C_{\alpha\beta} (C' \gamma^{a'})_{\alpha'\beta'} P_{a'} \bigg] \nonumber\\
&+ \epsilon_{IJ} \bigg[ C_{\alpha'\beta'} (C \gamma^{ab})_{\alpha\beta} J_{ab} - C_{\alpha\beta} (C' \gamma^{a'b'})_{\alpha'\beta'} J_{a'b'} \bigg]
\label{eq:superads}
\end{align}

\subsection{Super Poincaré Algebra}
\label{subsec:StringySuperPoincareAlgebra}
By applying the decomposition \autoref{3sub_expansion} to \autoref{eq:superads}, and by considering the lowest level expansion, we get the following $\mathfrak{psu}(2,2|4)(0,1, 2)$ superalgebra, which is a subalgebra of the $\mathcal{N}=2, D=10$ super Poincar\'e algebra. 
\begin{multicols}{2}
\setlength{\abovedisplayskip}{-15pt}
\allowdisplaybreaks
\begin{align}
[P_{a}, J_{bc}] &= 2\eta_{a[b} P_{c]} \nonumber\\
[Q_{I},J_{ab}]&=-\frac{1}{2}Q_{I}\gamma_{ab} \nonumber\\
[P_{a'}, J_{b'c'}] &= 2\eta_{a'[b'} P_{c']}  \nonumber\\
[Q_{I},J_{a'b'}]&=-\frac{1}{2}Q_{I}\gamma_{a'b'} \nonumber
\end{align}
\end{multicols}
\setlength{\parindent}{0pt}
\vspace{-0.9cm}
\begin{align}
[J_{ab} , J_{cd}] &= 2\eta_{c[b}J_{a]d}-2\eta_{d[b}J_{a]c} \nonumber\\ 
[J_{a'b'} , J_{c'd'}] &= 2\eta_{c'[b'}J_{a']d'} -2\eta_{d'[b'}J_{a']c'} \nonumber \\
\{ Q_{\alpha\alpha'I} , Q_{\beta\beta'J} \} &= \delta_{IJ} \bigg[ - 2i C_{\alpha'\beta'} (C \gamma^{a} )_{\alpha\beta} P_{a} + 2 C_{\alpha\beta} (C' \gamma^{a'})_{\alpha'\beta'} P_{a'} \bigg] 
\label{eq:superpoincare}
\end{align}

\subsection{Stringy Super-Newton-Hooke Algebra}
By applying the decomposition \autoref{NH_decomp} to \autoref{eq:superads}, and by considering the lowest level expansion, we get the following $\mathfrak{psu}(2,2|4)(0,1)$ superalgebra, which contains as a subalgebra the stringy Newton-Hooke superalgebra. 
\begin{multicols}{2}
\setlength{\abovedisplayskip}{-15pt}
\allowdisplaybreaks
\begin{align*}
[P_{\overline{a}} , P_{\overline{b}}] &= J_{\overline{a}\overline{b}} \nonumber\\
[P_{\overline{a}} , P_{\underline{b}}] &= J_{\overline{a}\underline{b}} \nonumber\\
[P_{\underline{a}}, J_{\underline{b}\underline{c}}] &= 2\eta_{\underline{a}[\underline{b}} P_{\underline{c}]}\nonumber\\
[P_{\overline{a}}, J_{\overline{b}\overline{c}}] &= 2\eta_{\overline{a}[\overline{b}} P_{\overline{c}]}\nonumber\\
[P_{\overline{a}}, J_{\overline{b}\underline{c}}] &= \eta_{\overline{a}\overline{b}} P_{\underline{c}}\nonumber\\
[P_{a'}, J_{b'c'}] &= 2\eta_{a'[b'} P_{c']} \nonumber\\
[J_{\overline{a}\overline{b}} , J_{\overline{c}\overline{d}}] &= 2\eta_{\overline{c}[\overline{b}}J_{\overline{a}]\overline{d}}-2\eta_{\overline{d}[\overline{b}}J_{\overline{a}]\overline{c}}\nonumber\\
[J_{\underline{a}\underline{b}} , J_{\underline{c}\underline{d}}] &= 2\eta_{\underline{c}[\underline{b}}J_{\underline{a}]\underline{d}}-2\eta_{\underline{d}[\underline{b}}J_{\underline{a}]\underline{c}}\nonumber\\
[J_{\overline{a}\overline{b}} , J_{\overline{c}\underline{d}}] &= 2\eta_{\overline{c}[\overline{b}}J_{\overline{a}]\underline{d}}\nonumber\\
[J_{\underline{a}\underline{b}} , J_{\overline{c}\underline{d}}] &= -2\eta_{\underline{d}[\underline{b}}J_{\underline{a}]\overline{c}}\nonumber\\
[J_{a'b'} , J_{c'd'}] &= 2\eta_{c'[b'}J_{a']d'} -2\eta_{d'[b'}J_{a']c'} \nonumber\\
[Q^{+} , P_{\underline{a}}] &=  \frac{1}{2}  Q^{-}\tau_{2} \gamma_{\underline{a}} \nonumber\\
[Q^{\pm} , P_{\overline{a}}] &=  \frac{1}{2}  Q^{\pm} \tau_{2}\gamma_{\overline{a}} \nonumber\\
[Q^{\pm},J_{\overline{a}\overline{b}}]&=-\frac{1}{2}Q^{\pm}\gamma_{\overline{a}\overline{b}}\nonumber\\
[Q^{\pm},J_{\underline{a}\underline{b}}]&=-\frac{1}{2}Q^{\pm}\gamma_{\underline{a}\underline{b}}\nonumber\\
[Q^{+},J_{\overline{a}\underline{b}}]&=-\frac{1}{2}Q^{-}\gamma_{\overline{a}\underline{b}}\nonumber\\
[Q^{+} , P_{a'}] &=  -\frac{i}{2}  Q^{-}\tau_{2} \gamma_{a'} \nonumber\\
[Q^{\pm},J_{a'b'}]&=-\frac{1}{2}Q^{\pm}\gamma_{a'b'}\nonumber\\
\end{align*}
\end{multicols}
\setlength{\parindent}{0pt}
\vspace{-0.9cm}
\begin{align}
\{ Q_{\alpha\alpha'}^{+} , Q_{\beta\beta'}^{+} \} &= \mathds{1} \bigg[ - 2i C_{\alpha'\beta'} (C \gamma^{\overline{a}} \Pi_{\pm})_{\alpha\beta} P_{\overline{a}} \bigg] \nonumber\\
&+ i\tau_{2} \bigg[ C_{\alpha'\beta'} (C \gamma^{
\overline{a}\overline{b}}\Pi_{\pm})_{\alpha\beta} J_{\overline{a}\overline{b}}+C_{\alpha'\beta'} (C \gamma^{\underline{a}\underline{b}}\Pi_{\pm})_{\alpha\beta} J_{\underline{a}\underline{b}}- (C\Pi_{\pm})_{\alpha\beta} (C' \gamma^{a'b'})_{\alpha'\beta'} J_{a'b'}  \bigg]\nonumber\\
\{ Q_{\alpha\alpha'}^{\pm} , Q_{\beta\beta'}^{\mp} \} &= \mathds{1} \bigg[ - 2i C_{\alpha'\beta'} (C \gamma^{\underline{a}}\Pi_{\mp} )_{\alpha\beta} P_{\underline{a}} + 2 (C\Pi_{\mp})_{\alpha\beta} (C' \gamma^{a'})_{\alpha'\beta'} P_{a'} \bigg] \nonumber\\
&+ i\tau_{2} \bigg[ 2C_{\alpha'\beta'} (C \gamma^{\overline{a}\underline{b}}\Pi_{\mp})_{\alpha\beta} J_{\overline{a}\underline{b}}  \bigg]
\end{align}\label{eq:superNH}

\subsection{Stringy Super-Galilei Algebra}
\label{subsec:StringySuperGalileiAlgebra}

By applying the decomposition \autoref{NH_decomp} to the algebra \autoref{eq:superpoincare}, or equivalently, by applying the decomposition \autoref{3sub_expansion} to the algebra \autoref{eq:superNH}, we obtain the following Super-Poincar\'e$(0,1)$, which contains as a subalgebra the stringy Galilei superalgebra. 
\begin{multicols}{2}
\setlength{\abovedisplayskip}{-15pt}
\allowdisplaybreaks
\begin{align*}
[P_{\underline{a}}, J_{\underline{b}\underline{c}}] &= 2\eta_{\underline{a}[\underline{b}} P_{\underline{c}]}\nonumber\\
[P_{\overline{a}}, J_{\overline{b}\overline{c}}] &= 2\eta_{\overline{a}[\overline{b}} P_{\overline{c}]}\nonumber\\
[P_{\overline{a}}, J_{\overline{b}\underline{c}}] &= \eta_{\overline{a}\overline{b}} P_{\underline{c}}\nonumber\\
[P_{a'}, J_{b'c'}] &= 2\eta_{a'[b'} P_{c']} \nonumber\\
[J_{\overline{a}\overline{b}} , J_{\overline{c}\overline{d}}] &= 2\eta_{\overline{c}[\overline{b}}J_{\overline{a}]\overline{d}}-2\eta_{\overline{d}[\overline{b}}J_{\overline{a}]\overline{c}}\nonumber\\
[J_{\underline{a}\underline{b}} , J_{\underline{c}\underline{d}}] &= 2\eta_{\underline{c}[\underline{b}}J_{\underline{a}]\underline{d}}-2\eta_{\underline{d}[\underline{b}}J_{\underline{a}]\underline{c}}\nonumber\\
[J_{\overline{a}\overline{b}} , J_{\overline{c}\underline{d}}] &= 2\eta_{\overline{c}[\overline{b}}J_{\overline{a}]\underline{d}}\nonumber\\
[J_{\underline{a}\underline{b}} , J_{\overline{c}\underline{d}}] &= -2\eta_{\underline{d}[\underline{b}}J_{\underline{a}]\overline{c}}\nonumber\\
[J_{a'b'} , J_{c'd'}] &= 2\eta_{c'[b'}J_{a']d'} -2\eta_{d'[b'}J_{a']c'} \nonumber\\
[Q^{\pm},J_{\overline{a}\overline{b}}]&=-\frac{1}{2}Q^{\pm}\gamma_{\overline{a}\overline{b}}\nonumber\\
[Q^{\pm},J_{\underline{a}\underline{b}}]&=-\frac{1}{2}Q^{\pm}\gamma_{\underline{a}\underline{b}}\nonumber\\
[Q^{+},J_{\overline{a}\underline{b}}]&=-\frac{1}{2}Q^{-}\gamma_{\overline{a}\underline{b}}\nonumber\\
[Q^{\pm},J_{a'b'}]&=-\frac{1}{2}Q^{\pm}\gamma_{a'b'}\nonumber\\
\end{align*}
\end{multicols}
\setlength{\parindent}{0pt}
\vspace{-0.9cm}
\begin{align}
\{ Q_{\alpha\alpha'}^{+} , Q_{\beta\beta'}^{+} \} &= \mathds{1} \bigg[ - 2i C_{\alpha'\beta'} (C \gamma^{\overline{a}} \Pi_{\pm})_{\alpha\beta} P_{\overline{a}} \bigg] \nonumber\\
\{ Q_{\alpha\alpha'}^{\pm} , Q_{\beta\beta'}^{\mp} \} &= \mathds{1} \bigg[ - 2i C_{\alpha'\beta'} (C \gamma^{\underline{a}}\Pi_{\mp} )_{\alpha\beta} P_{\underline{a}} + 2 (C\Pi_{\mp})_{\alpha\beta} (C' \gamma^{a'})_{\alpha'\beta'} P_{a'} \bigg] 
\end{align}\label{eq:superGalilei}

\subsection{BMN Algebra}
\label{subsec:BMNAlgebra}
In this section we report the pp-wave algebra commutation relations \cite{Hatsuda:2002xp}
\begin{multicols}{2}
\setlength{\abovedisplayskip}{-15pt}
\allowdisplaybreaks
\begin{align*}
[J_{\hat{i}\hat{j}},J_{\hat{k}\hat{l}}]&=2\eta_{\hat{k}[\hat{j}}J_{\hat{i}]\hat{l}}-2\eta_{\hat{l}[\hat{j}}J_{\hat{i}]\hat{k}}\nonumber\\
[P_{\hat{i}},J_{\hat{j}\hat{k}}]&=2\eta_{\hat{i}[\hat{j}}P_{\hat{k}]}\nonumber\\
[P^{*}_{\hat{i}},J_{\hat{j}\hat{k}}]&=2\eta_{\hat{i}[\hat{j}}P^{*}_{\hat{k}]}\nonumber\\
[P_{\hat{i}},P^{*}_{\hat{j}}]&=-\frac{1}{\sqrt{2}}\eta_{\hat{i}\hat{j}}P_{+}\nonumber\\
[P_{-},P_{\hat{i}}]&=-\frac{1}{\sqrt{2}}P^{*}_{\hat{i}}\nonumber\\
[P_{-},P^{*}_{\hat{i}}]&=\frac{1}{\sqrt{2}}P_{\hat{i}}\nonumber\\
[Q_{+,I},P_{-}]&=-\frac{1}{\sqrt{2}} Q_{+,J}I\epsilon_{IJ}\nonumber\\
[Q_{-,I},P_{\hat{i}}]&=-\frac{1}{2\sqrt{2}} Q_{+,J}\Gamma_{+}I\Gamma_{\hat{i}}\epsilon_{IJ}\nonumber\\
[Q_{-,I},P^{*}_{\hat{i}}]&=-\frac{1}{2\sqrt{2}}Q_{+,I}\Gamma_{+}\Gamma_{\hat{i}}\nonumber\\
[Q_{\pm,I},J_{\hat{i}\hat{j}}]&=-\frac{1}{2}Q_{\pm,I}\Gamma_{\hat{i}\hat{j}}\nonumber\\
\end{align*}
\end{multicols}
\setlength{\parindent}{0pt}
\vspace{-0.9cm}
\begin{align}
\{Q_{+,I},Q_{+,J}\}=&-2i\delta_{IJ}\mathcal{C}\Gamma^{+}P_{+}\nonumber\\
\{Q_{+,I},Q_{-,J}\}=&-i\delta_{IJ}\mathcal{C}\Gamma^{\hat{i}}P_{\hat{i}}\Gamma_{-}\Gamma_{+}-i\epsilon_{IJ}\mathcal{C}\Gamma_{-}\Gamma_{+}\big[\Gamma^{i}IP_{i}^{*}+\Gamma^{i'}JP_{i'}^{*}\big]\nonumber\\
\{Q_{-,I},Q_{-,J}\}=&-2i\delta_{IJ}\mathcal{C}\Gamma^{-}P_{-}
+\frac{i}{\sqrt{2}}\epsilon_{IJ}\mathcal{C}\Gamma^{-}\big[\Gamma^{ij}IJ_{ij}+\Gamma^{i'j'}JJ_{i'j'}\big]\ ,
\end{align}
where
\begin{align}
Q_{\pm,I}=\frac{1}{2}(1\pm \Gamma_{9}\Gamma_{0})Q_{I}
\end{align}
and 
\begin{align}
\Gamma_{\pm}&=\frac{1}{\sqrt{2}}(\Gamma_{9}\pm \Gamma_{0})\nonumber\\
I&=\Gamma_{1234}\nonumber\\
J&=\Gamma_{5678}\ .
\end{align}

\end{appendices}

\bibliography{LieAlgebraExpansionIntegrability20200528BJHEP}{}

\end{document}